\documentstyle[epsf,epsfig,aps,twocolumn,floats]{revtex}
\newcommand{\xx}{{\bf x}}

\renewcommand{\ss}{{\bf s}}
\newcommand{\Ss}{{\ss}}
\renewcommand{\aa}{{\bf a}}

\newcommand{\FF}{{\bf F}}
\newcommand{\JJ}{{\bf J}}
\newcommand{\uu}{{\bf u}}
\newcommand{\ff}{{\bf f}}

\newcommand{\be}{\begin{equation}}
\newcommand{\ee}{\end{equation}}
\newcommand{\bes}{\begin{eqnarray}}
\newcommand{\ees}{\end{eqnarray}}

\newcommand{\av}[1]{\langle #1 \rangle}

\newcommand{\tisean}{TISEAN}
\newcommand{\HOME}{$\tilde{~}$}
\newcommand{\THTTP}{{\tt http://www.mpipks-dresden.mpg.de/\HOME tisean}}
\newcommand{\Reals}{{\bf R}}

\newcommand{\NOISE}{{\tt noise}}
\newcommand{\CNAIVE}{{\tt c2naive}}
\newcommand{\MUTUAL}{{\tt mutual}}
\newcommand{\AUTOCOR}{{\tt autocor}}
\newcommand{\CORR}{{\tt corr}}
\newcommand{\PC}{{\tt pc}}
\newcommand{\POINCARE}{{\tt poincare}}
\newcommand{\SVD}{{\tt svd}}
\newcommand{\DELAY}{{\tt delay}}
\newcommand{\EMBED}{{\tt embed}}
\newcommand{\STP}{{\tt stp}}
\newcommand{\RECURR}{{\tt recurr}}
\newcommand{\GHKSS}{{\tt ghkss}}
\newcommand{\LLAR}{{\tt ll-ar}}
\newcommand{\PREDICT}{{\tt predict}}
\newcommand{\RBF}{{\tt rbf}}
\newcommand{\ZEROTH}{{\tt zeroth}}
\newcommand{\NSTEP}{{\tt nstep}}
\newcommand{\ONESTEP}{{\tt onestep}}
\newcommand{\POLYNOM}{{\tt polynom}}
\newcommand{\FNN}{{\tt false\_nearest}}
\newcommand{\EXTREMA}{{\tt extrema}}
\newcommand{\CONE}{{\tt c1}}
\newcommand{\CTWO}{{\tt c2}}
\newcommand{\CTOD}{{\tt c2d}}
\newcommand{\CTOG}{{\tt c2g}}
\newcommand{\CTOT}{{\tt c2t}}
\newcommand{\LYAPUNOV}{{\tt lyapunov}}
\newcommand{\LYAPSPEC}{{\tt lyap\_spec}}
\newcommand{\LYAPK}{{\tt lyap\_k}}
\newcommand{\BOXCOUNT}{{\tt boxcount}}
\newcommand{\SURROGATES}{{\tt surrogates}}
\newcommand{\LAZY}{{\tt lazy}}
\newcommand{\PROJECT}{{\tt project}}
\newcommand{\RGENER}{{\tt randomize\_generic\_exp\_random}}
\newcommand{\RAUTO}{{\tt randomize\_auto\_exp\_random}}

\newcommand{\UPO}{{\tt upo}}
\newcommand{\TIMEREV}{{\tt timerev}}
\newcommand{\AUTOCORTHREE}{{\tt autocor3}}
\newcommand{\DTWO}{{\tt d2}}
\newcommand{\LYAPR}{{\tt lyap\_r}}

\begin{document} 
\onecolumn

\title{Practical implementation of nonlinear time series methods: The
{\tisean} package}

\author{ 
{\large Rainer Hegger, Holger Kantz}\\[0.1cm]
   {\em Max Planck Institute for Physics of Complex Systems,\\ N\"othnitzer
     Str. 38, D--01187 Dresden}\\[0.3cm]
{\large Thomas Schreiber}\\[0.1cm]
   {\em Physics Department, University of Wuppertal, D--42097 Wuppertal}}
\maketitle
\begin{abstract} 
   We describe the implementation of methods of nonlinear time series analysis
   which are based on the paradigm of deterministic chaos. A variety of
   algorithms for data representation, prediction, noise reduction, dimension
   and Lyapunov estimation, and nonlinearity testing are discussed with
   particular emphasis on issues of implementation and choice of parameters.
   Computer programs that implement the resulting strategies are publicly
   available as the {\tisean} software package. The use of each algorithm will
   be illustrated with a typical application. As to the theoretical background,
   we will essentially give pointers to the literature.
\end{abstract}

\section*{Lead Paragraph}
\begin{quote}
Nonlinear time series analysis is becoming a more and more reliable tool for
the study of complicated dynamics from measurements. The concept of
low-dimensional chaos has proven to be fruitful in the understanding of many
complex phenomena despite the fact that very few natural systems have actually
been found to be low dimensional deterministic in the sense of the theory. In
order to evaluate the long term usefulness of the nonlinear time series
approach as inspired by chaos theory, it will be important that the
corresponding methods become more widely accessible.  This paper, while not a
proper review on nonlinear time series analysis, tries to make a contribution
to this process by describing the actual implementation of the algorithms, and
their proper usage. Most of the methods require the choice of certain
parameters for each specific time series application. We will try to give
guidance in this respect. The scope and selection of topics in this article, as
well as the implementational choices that have been made, correspond to the
contents of the software package {\tisean} which is publicly available from
\THTTP. In fact, this paper can be seen as an extended manual for the {\tisean}
programs. It fills the gap between the technical documentation and the existing
literature, providing the necessary entry points for a more thorough study of
the theoretical background.
\end{quote}
\twocolumn

\section{Introduction}
Deterministic chaos as a fundamental concept is by now well established and
described in a rich literature. The mere fact that simple deterministic systems
generically exhibit complicated temporal behavior in the presence of
nonlinearity has influenced thinking and intuition in many fields. However, it
has been questioned whether the relevance of chaos for the understanding of the
time evolving world goes beyond that of a purely philosophical paradigm.
Accordingly, major research efforts are dedicated to two related questions.
The first question is if chaos theory can be used to gain a better
understanding and interpretation of observed complex dynamical behavior. The
second is if chaos theory can give an advantage in predicting or controlling
such time evolution. Time evolution as a system property can be measured by
recording time series. Thus, nonlinear time series methods will be the key to
the answers of the above questions. This paper is intended to encourage the
explorative use of such methods by a section of the scientific community which
is not limited to chaos theorists. A range of algorithms has been made
available in the form of computer programs by the {\tisean}
project~\cite{tisean}. Since this is fairly new territory, unguided use of the
algorithms bears considerable risk of wrong interpretation and unintelligible
or spurious results. In the present paper, the essential ideas behind the
algorithms are summarized and pointers to the existing literature are given.
To avoid excessive redundancy with the text book~\cite{KantzSchreiber} and the
recent review~\cite{habil}, the derivation of the methods will be kept to a
minimum. On the other hand, the choices that have been made in the
implementation of the programs are discussed more thoroughly, even if this may
seem quite technical at times. We will also point to possible alternatives to
the {\tisean} implementation.

Let us at this point mention a number of general references on the subject of
nonlinear dynamics.  At an introductory level, the book by Kaplan and
Glass~\cite{KaplanGlass} is aimed at an interdisciplinary audience and provides
a good intuitive understanding of the fundamentals of dynamics.  The
theoretical framework is thoroughly described by Ott~\cite{Ott}, but also in
the older books by Berg\'e et al.~\cite{Berge} and by
Schuster~\cite{Schuster}. More advanced material is contained in the work by
Katok and Hasselblatt~\cite{KatokHasselblatt}. A collection of research
articles compiled by Ott et al.~\cite{coping} covers some of the more applied
aspects of chaos, like synchronization, control, and time series analysis.

Nonlinear time series analysis based on this theoretical paradigm is described
in two recent monographs, one by Abarbanel~\cite{abarbook} and one by Kantz and
Schreiber~\cite{KantzSchreiber}. While the former volume usually {\em assumes}
chaoticity, the latter book puts some emphasis on practical applications to
time series that are not manifestly found, nor simply assumed to be,
deterministic chaotic. This is the rationale we will also adopt in the present
paper. A number of older articles can be seen as reviews, including Grassberger
et al.~\cite{gss}, Abarbanel et al.~\cite{abarbanel}, as well as Kugiumtzis et
al.~\cite{kugiumtzis_rev1,kugiumtzis_rev2}.  The application of nonlinear time
series analysis to real world measurements where determinism is unlikely to be
present in a stronger sense, is reviewed in Schreiber~\cite{habil}.  Apart from
these works, a number of conference proceedings volumes are devoted to chaotic
time series, including Refs.~\cite{Mayer-Kress,casdagli,SFI,dyndis,freital}.

\subsection{Philosophy of the {\tisean} implementation}\label{sec:philo}
A number of different people have been credited for the saying that every
complicated question has a simple answer which is wrong. Analysing a time
series with a nonlinear approach is definitely a complicated problem. Simple
answers have been repeatedly offered in the literature, quoting numerical
values for attractor dimensions for any conceivable system.  The present
implementation reflects our scepticism against such simple answers which are
the inevitable result of using black box algorithms.  Thus, for example, none
of the ``dimension'' programs will actually print a number which can be quoted
as the estimated attractor dimension. Instead, the correlation sum is computed
and basic tools are provided for its interpretation. It is up to the scientist
who does the analysis to put these results into their proper context and to
infer what information she or he may find useful and plausible. We should
stress that this is not simply a question of error bars. Error bars don't tell
about systematic errors and neither do they tell if the underlying assumptions
are justified.

The {\tisean} project has emerged from work of the involved research groups
over several years. Some of the programs are in fact based on code published in
Ref.~\cite{KantzSchreiber}. Nevertheless, we still like to see it as a starting
point rather than a conclusive step. First of all, nonlinear time series
analysis is still a rapidly evolving field, in particular with respect to
applications. This implies that the selection of topics in this article and the
selection of algorithms implemented in {\tisean} are highly biased towards what
we know now and found useful so far. But even the well established concepts
like dimension estimation and noise reduction leave considerable room for
alternatives to the present implementation. Sometimes this resulted in two or
more concurring and almost redundant programs entering the package. We have
deliberately not eliminated these redundancies since the user may benefit from
having a choice.  In any case it is healthy to know that for most of the
algorithms the final word hasn't been spoken yet -- nor is ever to be.

While the {\tisean} package does contain a number of tools for {\em linear}
time series analysis (spectrum, autocorrelations, histograms, etc.), these are
only suitable for a quick inspection of the data. Spectral or even ARMA
estimation are industries in themselves and we refer the reader -- and the user
of {\tisean} -- to the existing literature and available statistics software
for optimal, up-to-date implementations of these important methods.

Some users will miss a convenient graphical interface to the programs. We felt
that at this point the extra implementational effort would not be justified by
the expected additional functionality of the package. Work is in progress,
however, to provide interfaces to higher level mathematics (or statistics)
software.

\subsection{General computational issues}\label{sec:general}
The natural basis to formulate nonlinear time series algorithms from chaos
theory is a multi-dimensional phase space, rather than the time or the
frequency domain. It will be essential for the global dynamics in this phase
space to be nonlinear in order to fulfill the constraints of non-triviality and
boundedness. Only in particular cases this nonlinear structure will be easily
representable by a global nonlinear function. Instead, all properties will be
expressed in terms of local quantities, often by suitable global averages.  All
local information will be gained from neighborhood relations of various kinds
from time series elements.  Thus a recurrent computational issue will be that
of defining local neighborhoods in phase space. Finding neighbors in
multidimensional space is a common problem of computational
geometry. Multidimensional tree structures are widely used and have attractive
theoretical properties. Finding all neighbors in a set of $N$ vectors takes
$O(\log N)$ operations, thus the total operation count is $O(N\log N)$.  A fast
alternative that is particularly efficient for relatively low dimensional
structures embedded in multidimensional spaces is given by box-assisted
neighbor search methods which can push the operation count down to $O(N)$
under certain assumptions. Both approaches are reviewed in Ref.~\cite{neigh}
with particular emphasis on time series applications. In the {\tisean} project,
fast neighbor search is done using a box-assisted approach, as described in
Ref.~\cite{KantzSchreiber}.

No matter in what space dimension we are working, we will define {\em
candidates} for nearest neighbors in two dimensions using a grid of evenly
spaced boxes. With a grid of spacing $\epsilon$, all neighbors of a vector
$\bf x$ closer than epsilon must be located in the adjacent boxes. But not all
points in the adjacent boxes are neighbors, they may be up to $2\epsilon$
away in two dimensions and arbitrarily far in higher dimensions. Neighbors
search is thus a two stage process. First the box-assisted data base has to be
filled and then for each point a list of neighbors can be requested. 
There are a few instances where it is advisable to abandon the fast neighbor
search strategy. One example is the program {\NOISE} that does nonlinear noise
filtering in a data stream. It is supposed to start filtering soon after the
first points have been recorded. Thus the neighbor data base cannot be
constructed in the beginning. Another exception is if quite short ($<500$
points, say), high dimensional data are processed. Then the overhead for the
neighbor search should be avoided and instead an optimized straight $O(N^2)$
method be used, like it is done in~{\CNAIVE}.

For portability, all programs expect time series data in column format
represented by ASCII numbers. The column to be processed can be specified on
the command line. Although somewhat wasteful for storing data, ASCII is the
least common divisor between the different ways most software can store data.
All parameters can be set by adding options to the command, which, in many
programs, just replace the default values. Obviously, relying on default
settings is particularly dangerous in such a subtle field.  Since almost all
routines can read from standard input and write to standard output, programs
can be part of pipelines. For example they can be called as filters from inside
graphics software or other software tools which are able to execute shell
commands.  Also, data conversion or compression can be done ``on the fly'' this
way. The reader here realizes that we are speaking of UNIX or LINUX platforms
which seems to be the most appropriate environment. It is however expected that
most of the programs will be ported to other environments in the near future.

For those readers familiar with the programs published in
Ref.~\cite{KantzSchreiber} we should mention that these form the basis for a
number of those {\tisean} programs written in FORTRAN. The C programs, even if
they do similar things, are fairly independent implementations. All C and C++
programs now use dynamic allocation of storage, for example.

\begin{figure}[t]
\begin{center}
   ~\\[-15pt]
   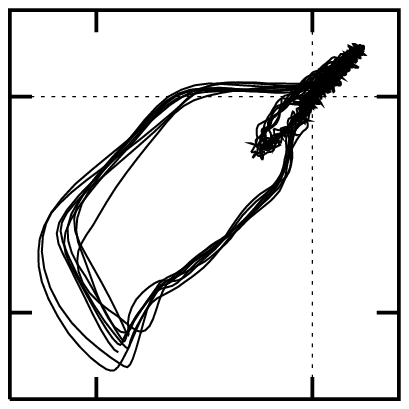\\[-5pt]
   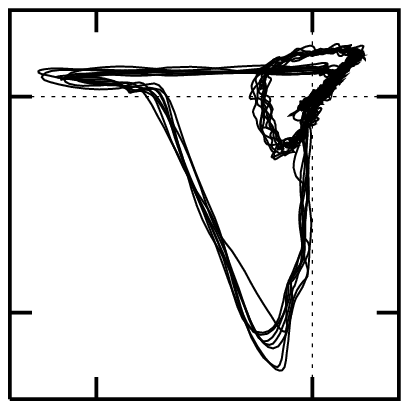
\end{center}
\caption[]{\label{fig:mcgd}\small
   Time delay representation of a human magneto-cardiogram. In the upper panel,
   a short delay time of 10~ms is used to resolve the fast waveform
   corresponding to the contraction of the ventricle. In the lower panel, the
   slower recovery phase of the ventricle (small loop) is better resolved due
   to the use of a slightly longer delay of 40~ms.  Such a plot can be
   conveniently be produced by a graphic tool such as {\tt gnuplot} without
   generating extra data files.}
\end{figure}

\section{Phase space representation}\label{sec:delay}
Deterministic dynamical systems describe the time evolution of a system in some
phase space $\Gamma\subset\Reals^d$.  They can be expressed for example by
ordinary differential equations
\be\label{eq:ode}
   \dot\xx(t)=\FF(\xx(t))
\,,\ee
or in discrete time $t=n\Delta t$ by maps of the form
\be\label{eq:map}
   \xx_{n+1}=\ff(\xx_n)
\,.\ee
A time series can then be thought of as a sequence of observations
$\{s_n=s(\xx_n)\}$ performed with some measurement function $s(\cdot)$.  Since
the (usually scalar) sequence $\{s_n\}$ in itself does not properly represent
the (multidimensional) phase space of the dynamical system, one has to employ some
technique to unfold the multidimensional structure using the available data.

\subsection{Delay coordinates}
The most important phase space reconstruction technique is the {\em method of
delays}. Vectors in a new space, the embedding space, are formed from time
delayed values of the scalar measurements:
\be\label{eq:delay}
   \Ss_n=(s_{n-(m-1)\tau},s_{n-(m-2)\tau}, \ldots, s_n)
\,.\ee
The number $m$ of elements is called the {\em embedding dimension}, the time
$\tau$ is generally referred to as the {\em delay} or {\em lag}.  Celebrated
embedding theorems by Takens~\cite{takens} and by Sauer et al.~\cite{embed}
state that if the sequence $\{s_n\}$ does indeed consist of scalar measurements
of the state of a dynamical system, then under certain genericity assumptions,
the time delay embedding provides a one-to-one image of the original set
$\{\xx\}$, provided $m$ is large enough.

Time delay embeddings are used in almost all methods described in this
paper. The implementation is straightforward and does not require further
explanation.  If $N$ scalar measurements are available, the number of embedding
vectors is only $N-(m-1)\tau$. This has to be kept in mind for the correct
normalization of averaged quantities. There is a large literature on the
``optimal'' choice of the embedding parameters $m$ and $\tau$. It turns out,
however, that what constitutes the optimal choice largely depends on the
application. We will therefore discuss the choice of embedding parameters
occasionally together with other algorithms below.

A stand-alone version of the delay procedure ({\DELAY}, {\EMBED}) is an
important tool for the visual inspection of data, even though visualization is
restricted to two dimensions, or at most two-dimensional projections of
three-dimensional renderings. A good unfolding already in two dimensions may
give some guidance about a good choice of the delay time for higher
dimensional embeddings.  As an example let us show two different
two-dimensional delay coordinate representations of a human magneto-cardiogram
(Fig.~\ref{fig:mcgd}). Note that we do neither assume nor claim that the
magneto- (or electro-) cardiogram is deterministic or even chaotic. Although
in the particular case of cardiac recordings the use of time delay embeddings
can be motivated theoretically~\cite{marcus}, we here only want to use the
embedding technique as a visualization tool.

\subsection{Embedding parameters}
A reasonable choice of the delay gains importance through the fact that we
always have to deal with a finite amount of noisy data. Both noise and
finiteness will prevent us from having access to infinitesimal length scales,
so that the structure we want to exploit should persists up to the largest
possible length scales. Depending on the type of structure we want to explore
we have to choose a suitable time delay. Most obviously, delay unity for
highly sampled flow data will yield delay vectors which are all concentrated
around the diagonal in the embedding space and thus all structure
perpendicular to the diagonal is almost invisible. In~\cite{Casdagli} the
terms {\sl redundancy} and {\sl irrelevance} were used to characterize the
problem: Small delays yield strongly correlated vector elements, large delays
lead to vectors whose components are (almost) uncorrelated and the data are
thus (seemingly) randomly distributed in the embedding space. Quite a number
of papers have been published on the proper choice of the delay and embedding
dimension. We have argued repeatedly~\cite{gss,KantzSchreiber,habil} that an
``optimal'' embedding can -- if at all -- only be defined relative to a
specific purpose for which the embedding is used. Nevertheless, some
quantitative tools are available to guide the choice.

The usual autocorrelation function ({\AUTOCOR}, {\CORR}) and the time delayed
mutual information ({\MUTUAL}), as well as visual inspection of delay
representations with 
various lags provide important information about reasonable delay times while
the false neighbors statistic ({\FNN}) 
can give guidance about the proper embedding
dimension. Again, ``optimal'' parameters cannot be thus established except in
the context of a specific application. 

\subsubsection{Mutual information}
The time delayed mutual information was suggested by Fraser and
Swinney~\cite{fraser} as a tool to determine a reasonable delay: Unlike the
autocorrelation function, the mutual information takes into account also
nonlinear correlations. One has to compute
\be\label{eq:delay.mutual} 
   S=- \sum_{ij} p_{ij}(\tau)\ln\frac{p_{ij}(\tau)}{p_ip_j}
\,,\ee 
where for some partition on the real numbers $p_i$ is the probability to find
a time series value in the $i$-th interval, and $p_{ij}(\tau)$ is the joint
probability that an observation falls into the $i$-th interval and the
observation time $\tau$ later falls into the $j$-th. In theory this expression
has no systematic dependence on the size of the partition elements and can be
quite easily computed. There exist good arguments that if the time delayed
mutual information exhibits a marked minimum at a certain value of $\tau$,
then this is a good candidate for a reasonable time delay. However, these
arguments have to be modified when the embedding dimension exceeds two.
Moreover, as will become transparent in the following sections, not all
applications work optimally with the same delay. Our routine {\MUTUAL} uses
Eq.(\ref{eq:delay.mutual}), where the number of boxes of identical size and
the maximal delay time has to be supplied. The adaptive algorithm used
in~\cite{fraser} is more data intensive. Since we are not really interested in
absolute values of the mutual information here but rather in the first
minimum, the minimal implementation given here seems to be sufficient. The
related generalized mutual information of order two can be defined using the
correlation sum concept (Sec.\ref{sec:dimension}, \cite{pompe,milan}).
Estimation of the correlation entropy is explained in Sec.\ref{sec:dim.c2}.

\begin{figure}[t]
\centerline{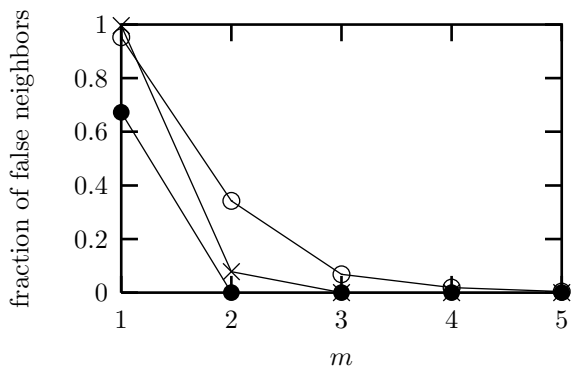}
\caption[]{\label{fig:fnn}\small 
   The fraction of false nearest neighbors as a function of the embedding
   dimension for noise free Lorenz (crosses) and H\'enon (filled circles)
   time series, as well as a H\'enon time series (open circles) corrupted by
   10\% of noise.}
\end{figure}

\subsubsection{False nearest neighbors}
A method to determine the minimal sufficient embedding dimension $m$ was
proposed by Kennel et al.~\cite{kennelFNN}. It is called the {\it false nearest
neighbor} method. The idea is quite intuitive. Suppose the minimal embedding
dimension for a given time series $\{s_i\}$ is $m_0$. This means that in a
$m_0$-dimensional delay space the reconstructed attractor is a one-to-one image
of the attractor in the original phase space. Especially, the topological
properties are preserved. Thus the neighbors of a given point are mapped onto
neighbors in the delay space. Due to the assumed smoothness of the dynamics,
neighborhoods of the points are mapped onto neighborhoods again. Of course the
shape and the diameter of the neighborhoods is changed according to the
Lyapunov exponents.  But suppose now you embed in an $m$-dimensional space with
$m<m_0$. Due to this projection the topological structure is no longer
preserved. Points are projected into neighborhoods of other points to which
they wouldn't belong in higher dimensions. These points are called {\it false
neighbors}. If now the dynamics is applied, these false neighbors are not
typically mapped into the image of the neighborhood, but somewhere else, so
that the average ``diameter'' becomes quite large.

The idea of the algorithm {\FNN} is the following. For each point $\vec{s}_i$
in the time series look for its nearest neighbor $\vec{s}_j$ in a
$m$-dimensional space. Calculate the distance $\|\vec{s}_i-\vec{s}_j\|$.
Iterate both points and compute
\be
   R_i=\frac{|s_{i+1}-s_{j+1}|}{\|\vec{s}_i-\vec{s}_j\|}
\,.\ee
If $R_i$ exceeds a given heuristic threshold $R_t$, this point is marked as
having a false nearest neighbor~\cite{kennelFNN}.  The criterion that the
embedding dimension is high enough is that the fraction of points for which
$R_i>R_t$ is zero, or at least sufficiently small. Two examples are shown in
Fig.~\ref{fig:fnn}. One is for the Lorenz system (crosses), one for the H\'enon
system (filled circles), and one for a H\'enon time series corrupted by 10\% of
Gaussian white noise (open circles). One clearly sees that, as expected, $m=2$
is sufficient for the H\'enon and $m=3$ for the Lorenz system, whereas the
signature is less clear in the noisy case.

The introduction of the false nearest neighbors concept and other ad hoc
instruments was partly a reaction to the finding that many results obtained for
the genuine invariants, like the correlation dimension, has been spurious
due to caveats of the estimation procedure. In the latter case, serial
correlations and small sample fluctuations can easily be mistaken for  
nonlinear determinism. It turns out, however, that the ad hoc quantities
basically suffer from the same problems - which can be cured by the same
precautions. The implementation {\FNN} therefore allows to specify a minimal
temporal separation of valid neighbors.

Other software for the analysis of false nearest neighbors is available in
source form from Kennel~\cite{kennel}. Or, if you prefer to pay for a license,
from Ref.~\cite{abla}.

\begin{figure}
\centerline{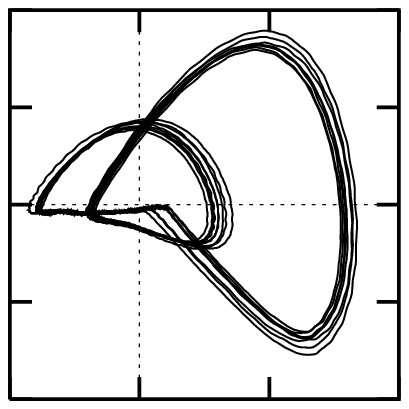}
\caption[]{\label{fig:mcg_pc}\small
   Phase space representation of a human magneto-cardiogram using the two
   largest principal components. An initial embedding was chosen in $m=20$
   dimensions with a delay of $\tau=7$~ms. The two components cover 70\% of the
   variance of the initial embedding vectors.}
\end{figure}

\subsection{Principal components}\label{sec:svd}
It has been shown in Ref.~\cite{embed} that the embedding technique can be
generalized to a wide class of smooth transformations applied to a time delay
embedding. In particular, if we introduce time delay coordinates $\{\Ss_n\}$,
then almost every linear transformation of sufficient rank again leads to an
embedding. A specific choice of linear transformation is known as {\em
principal component analysis, singular value decomposition, empirical
orthogonal functions, Karhunen-Lo\'eve decomposition}, and probably a few other
names. The technique is fairly widely used, for example to reduce multivariate
data to a few major modes. There is a large literature, including textbooks
like that by Jolliffe~\cite{PC}. In the context of nonlinear signal processing,
the technique has been advocated among others by Broomhead and King~\cite{svd}.

The idea is to introduce a new set of orthonormal basis vectors in embedding
space such that projections onto a given number of these directions preserve
the maximal fraction of the variance of the original vectors. In other words,
the error in making the projection is minimized for a given number of
directions. Solving this minimization problem~\cite{PC} leads to an eigenvalue
problem. The desired {\em principal directions} can be obtained as the
eigenvectors of the symmetric autocovariance matrix that correspond to the
largest eigenvalues. The alternative and formally equivalent approach via the
trajectory matrix is used in Ref.~\cite{svd}. The latter is numerically more
stable but involves the singular value decomposition of an $N\times m$ matrix
for $N$ data points embedded in $m$ dimensions, which can easily exceed
computational resources for time series of even moderate length~\cite{numrec}.

In almost all the algorithms described below, simple time delay embeddings can
be substituted by principal components. In the {\tisean} project (routines
{\SVD}, {\PC}), principal components are only provided as a stand-alone
visualization tool and for linear filtering~\cite{Vautard}, see
Sec.~\ref{sec:svdfilter} below.  In any case, one first has to choose an
initial time delay embedding and then a number of principal components to be
kept. For the purpose of visualization, the latter is immediately restricted
to two or at most three.  In order to take advantage of the noise averaging
effect of the principal component scheme, it is advisable to choose a much
shorter delay than one would for an ordinary time delay embedding, while at
the same time increasing the embedding dimension.  Experimentation is
recommended.  Figure~\ref{fig:mcg_pc} shows the contributions of the first two
principal components to the magneto-cardiogram shown in Fig.~\ref{fig:mcgd}.

\subsection{Poincar\'e sections}
Highly sampled data representing the continuous time of a differential
equation are called {\em flow} data. They are characterized by the fact that
errors in the direction tangent to the trajectory do neither shrink nor
increase exponentially (so called marginally stable direction) and thus
possess one Lyapunov exponent which is zero, since any perturbation in this
direction can be compensated by a simple shift of the time. Since in many data
analysis tasks this direction is of low interest, one might wish to eliminate
it. The theoretical concept to do so is called the Poincar\'e section. After
having chosen an $(m-1)$-dimensional hyperplane in the $m$-dimensional
(embedding) space, one creates a compressed time series of only the
intersections of the time continuous trajectory with this hyperplane {\sl in a
  predefined orientation}. These data are then vector valued discrete time
{\sl map like} data.  One can consider the projection of these
$(m-1)$-dimensional vectors onto the real numbers as another measurement
function (e.g.\ by recording the value of $s_n$ when $\ss_n$ passes the
Poincar\'e surface), so that one can create a new scalar time series if
desirable. The program {\POINCARE} constructs a sequence of vectors from a
scalar flow-like data set, if one specifies the hyperplane, the orientation,
and the embedding parameters. The intersections of the discretely sampled
trajectory with the Poincar\'e plane are computed by a third order
interpolation.

\begin{figure}[t]
\begin{center}
   ~\\[-15pt]
   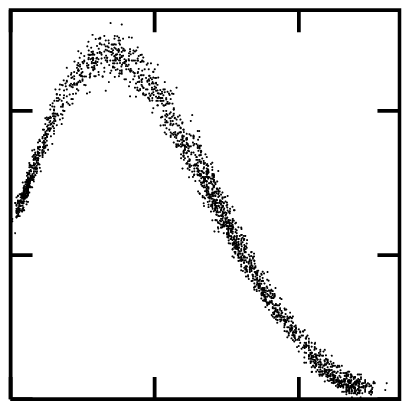\\[-5pt]
   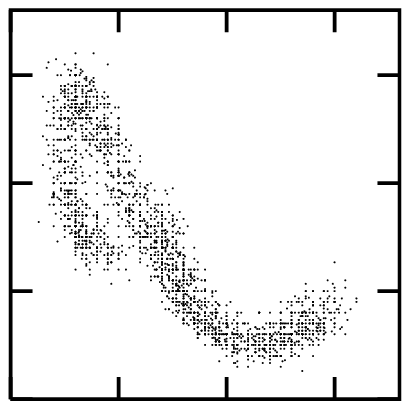
\end{center}
\caption[]{\label{fig:delay.Poincare}\small 
   Poincar\'e surface of section using {\EXTREMA}: A two-dimensional delay
   plot of the sequence of maxima (top) and of the time intervals between
   successive maxima (bottom). 
   without employing the
   option {\tt -t} {\sl time}, where {\sl time} is the number of time steps
   after the last extremum during which no further extrema are searched for
   (here: 3), one finds some fake extrema due to noise showing up close to the
   diagonal of the delay representation.
   Data: Time series of the output power of a CO$_2$ 
   laser~\cite{INO}. }
\end{figure}

The placement of the Poincar\'e surface is of high relevance for the usefulness
of the result. An optimal surface maximizes the number of intersections, i.e.\
minimizes the time intervals between them, if at the same time the attractor
remains connected. One avoids the trials and errors related to that if one
defines a surface by the zero crossing of the temporal derivative of the
signal, which is synonymous with collecting all maxima or all minima,
respectively.  This is done by \EXTREMA. However, this method suffers more from
noise, since for small time derivatives (i.e.\ close to the extrema) additional
extrema can be produced by perturbations.  Another aspect for the choice of the
surface of section is that one should try to maximize the variance of the data
inside the section, since their absolute noise level is independent of the
section. One last remark: Time intervals between intersections are phase space
observables as well~\cite{Hegger+} and the embedding theorems are thus
valid. For time series with pronounced spikes, one often likes to study the
sequence of interspike time intervals, e.g.\ in cardiology the RR-intervals. If
these time intervals are constructed in a way to yield time intervals of a
Poincar\'e map, they are suited to reflect the deterministic structure (if
any). For complications see~\cite{Hegger+}.

For a periodically driven non-autonomous system the best surface of section is
usually given by a fixed phase of the driving term, which is also called a
{\sl stroboscopic view}. Here again the selection of the phase should be
guided by the variance of the signal inside the section.

\subsection{SVD filters}\label{sec:svdfilter}
There are at least two reasons to apply an SVD filter to time series data:
Either, if one is working with flow data, one can implicitly determine the
optimal time delay, or, when deriving a stroboscopic map from synchronously
sampled data of a periodically driven system, one might use the redundancy to
optimize the signal to noise ratio.

In both applications the mathematics is the same: One constructs the
covariance matrix of all data vectors (e.g.\ in an $m$-dimensional
time delay embedding space), 
\be
   C_{ij}=\av{s_{n-m+i}s_{n-m+j}} - \av{s_{n-m+i}}\av{s_{n-m+j}}
\,,\ee
and computes its singular vectors. Then one projects onto the
$m$-dimensional vectors
corresponding to the $q$ largest singular values. To work with flow data, $q$
should be at least the correct embedding dimension, and $m$ considerably 
larger (e.g.\ $m=2q$ or larger). The result is a vector valued time series, and
in~\cite{embed} the relation of these components to temporal derivatives on
the one hand and to Fourier components on the other hand were discussed. If,
in the non-autonomous case, one wants to compress flow data to map data,
$q=1$. In this case, the redundancy of the flow is implicitly used for noise 
reduction of the map data. The routine {\SVD} can be used for both purposes.

\section{Visualization, non-stationarity}\label{sec:visual}
\subsection{Recurrence plots}
\begin{figure}
\centerline{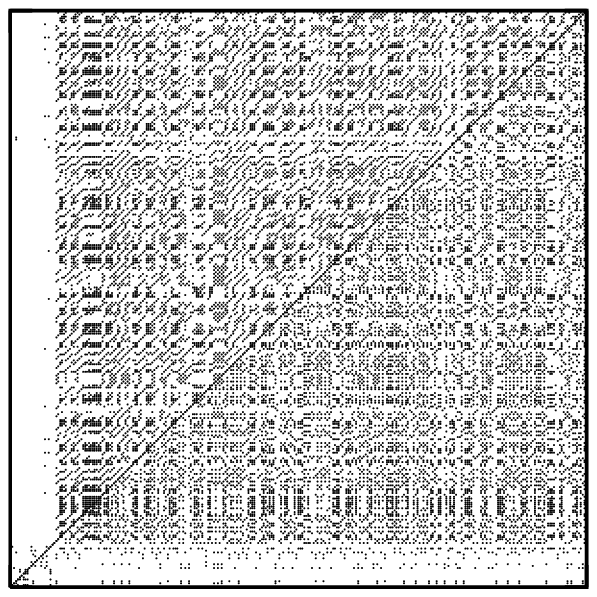}
\caption[]{\label{fig:visual.recur}\small 
   Recurrence plot for Poincar\'e section data from a vibrating string
   experiment~\cite{string}. Above the diagonal an embedding in two dimensions
   was used while below the diagonal, scalar time series values were compared.
   In both cases the lighter shaded region at the beginning of the recording
   indicates that these data are dynamically distinct from the rest. In this
   particular case this was due to adjustments in the measurement apparatus.}
\end{figure}

Recurrence plots are a useful tool to identify structure in a data set in a
time resolved way qualitatively. This can be intermittency (which one detects
also by direct inspection), the temporary vicinity of a chaotic trajectory to
an unstable periodic orbit, or non-stationarity.  They were introduced
in~\cite{Ruelle} and investigated in much detail in \cite{Casdagli_recurr},
where you find many hints on how to interprete the results. Our routine
{\RECURR} simply scans the time series and marks each pair of time indices
$(i,j)$ with a black dot, whose corresponding pair of delay vectors has
distance $\le \epsilon$. Thus in the $(i,j)$-plane, black dots indicate
closeness. In an ergodic situation, the dots should cover the plane uniformly
on average, whereas non-stationarity expresses itself by an overall tendency
of the dots to be close to the diagonal. Of course, a return to a dynamical
situation the system was in before becomes evident by a black region far away
from the diagonal. In Fig.~\ref{fig:visual.recur}, a recurrence plot is used
to detect transient behavior at the beginning of a longer recording.

For the purpose of stationary testing, the recurrence plot is not particularly
sensitive to the choice of embedding. The contrast of the resulting images can
be selected by the distance $\epsilon$ and the percentage of dots that should
be actually plotted. Various software involving the color rendering and
quantification of recurrence plots is offered in DOS executable form by
Webber~\cite{webber}. The interpretation of the often intriguing patterns
beyond the detection and study of non-stationarity is still an open question.
For suggestions for the study of nonstationary signals see~\cite{habil} and
references given there.

\begin{figure}[t]
\centerline{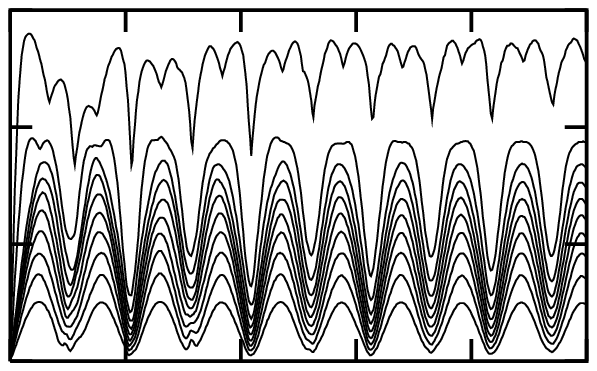}
\caption[]{\label{fig:visual.stp}\small 
   Space-time separation plot of the CO$_2$ laser data. Shown are lines of
   constant probability density of a point to be $\epsilon$-neighbor of the
   current point if its temporal distance is $\delta t$. Probability densitites
   are 1/10 to 1 with increments of 1/10 from bottom to top.  Clear
   correlations are visible.} 
\end{figure}

\subsection{Space-time separation plot}\label{sec:stp}
While the recurrence plot shows absolute times, the space-time separation plot
introduced by Provenzale et al.~\cite{stp} integrates along parallels to the
diagonal and thus only shows relative times. One usually draws lines of
constant probability per time unit of a point to be an $\epsilon$-neighbor of
the current point, when its time distance is $\delta t$. This helps
identifying temporal correlations inside the time series and is relevant to
estimate a reasonable delay time, and, more importantly, the Theiler-window $w$
in dimension and Lyapunov-analysis (see Sec.~\ref{sec:dimension}). Said in
different words, it shows how large the temporal distance between points should
be so that we can assume that they form independent samples according to the
invariant measure. The corresponding routine of the {\tisean} package
is~{\STP}, see Fig.~\ref{fig:visual.stp}.

\section{Nonlinear prediction}\label{sec:predict}
To think about predictability in time series data is worth while even if one
is not interested in forecasts at all. Predictability is one way how
correlations between data express themselves. These can be linear
correlations, nonlinear correlations, or even deterministic contraints.
Questions related to those relevant for predictions will reappear with noise
reduction and in surrogate data tests, but also for the computation of
Lyapunov exponents from data.  Prediction is discussed in most of the general
nonlinear time series references, in particular, a nice collection of articles
can be found in~\cite{SFI}.

\subsection{Model validation}
Before entering the methods, we have to discuss how to assess the results. The
most obvious quantity for the quantification of predictability is the average
forecast error, i.e.\ the root of the mean squared (rms) deviation of the
individual prediction from the actual future value. If it is computed on those
values which were also used to construct the model (or to perform the
predictions), it is called the {\sl in-sample error}. It is always advisable
to save some data for an out-of-sample test. If the out-of-sample error is
considerably larger than the in-sample error, data are either non-stationary
or one has overfitted the data, i.e.\ the fit extracted structure from random
fluctuations. A model with less parameters will then serve better.  In cases
where the data base is poor, on can apply {\sl complete cross-validation} or
{\sl take-one-out statistics}, i.e.\ one constructs as many models as one
performs forecasts, and in each case ignores the point one wants to predict.
By construction, this method is realized in the local approaches, but not in
the global ones.

The most significant, but least quantitative way of model validation is to
iterate the model and to compare this synthetic time series to the
experimental data. If they are compatible (e.g.\ in a delay plot), then the
model is likely to be reasonable. Quantitatively, it is not easy to define the
compatibility. One starts form an observed delay vector as intial condition,
performs the first forecast, combines the forecast with all but the last
components of the initial vector to a new delay vector, performs the next
forecast, and so on. The resulting time series should then be compared to the
measured data, most easily the attractor in a delay representation.

\begin{figure}
\centerline{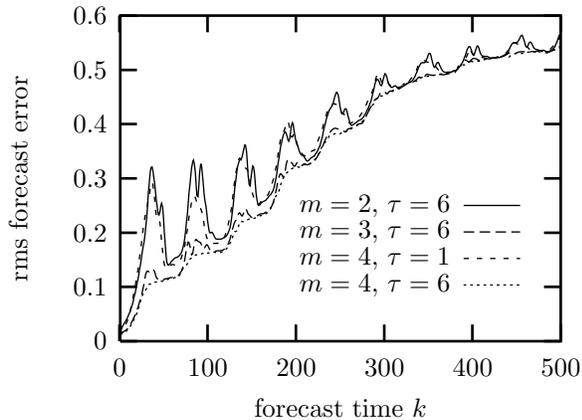}
\caption[]{\label{fig:predict.INOzero}\small 
  Predictions $k$ time steps ahead (no iterated predictions) using the program
  {\ZEROTH}. Top curve: embedding dimension two is insufficient, since these
  flow data fill a (2+$\epsilon$)-dimensional attractor.  Second from top:
  Although embedding dimension four should in theory be a good
  embedding, $\tau=1$ suppresses structure perpendicular to the diagonal so
  that the predictions are as bad as in $m=2$! Lower curves: $m=3$ and $4$ with
  a delay of about 4-8 time units serve well. }
\end{figure}

\subsection{Simple nonlinear prediction}\label{sec:zeroth}
Conventional linear prediction schemes average over all locations in phase
space when they extract the correlations they exploit for predictability.
Tong~\cite{TAR} promoted an extension that fits different linear models if the
current state is below or above a given threshold (TAR, {\bf T}hreshold {\bf
A}utoregressive {\bf M}odel). If we expect more than a slight nonlinear
component to be present, it is preferable to make the approximation as local in
phase space as possible.  There have been many similar suggestions in the
literature how to exploit local structure, see
e.g.~\cite{Arkady,sugimay,Eckmann,fsid0}. The simplest approach is to make the
approximation local but only keep the zeroth order, that is, approximate the
dynamics locally by a constant.  In the {\tisean} package we include such a
robust and simple method: In a delay embedding space, all neighbors of $\ss_n$
are saught, if we want to predict the measurements at time $n+k$. The forecast
is then simply
\be\label{eq:predict.lazy}
   \hat s_{n+k}= {1\over |{\cal U}_n|} \sum_{\ss_j\in{\cal U}_n} s_{j+k}
\,,\ee
i.e.\ the average over the ``futures'' of the neighbors. The average forecast
errors obtained with the routine {\ZEROTH} ({\PREDICT} would give similar
results) for the laser output data used in Fig.~\ref{fig:delay.Poincare} as a
function of the number $k$ of steps ahead the predictions are made is shown in
Fig.~\ref{fig:predict.INOzero}. One can also iterate the predictions by using
the time series as a data base.

Apart from the embedding parameters, all that has to be specified for zeroth
order predictions is the size of the neighborhoods. Since the diffusive
motion below the noise level cannot be predicted anyway, it makes sense to
select neighborhoods which are at least as large as the noise level, maybe
two or three times larger. For fairly clean time series, this guideline may
result in neighborhoods with very few points. Therefore {\ZEROTH} also
permits to specify the minimal number of neighbors to base the predictions
on.

A relevant modification of this method is to extend the neighborhood ${\cal
U}$ to infinity, but to introduce a distance dependent weight,
\be\label{eq:predict.kernel}
   \hat s_{n+k}= {\sum_{j\ne n} s_{j+k} w(|\ss_n-\ss_j|) 
                 \over  \sum_{j\ne n} w(|\ss_n-\ss_j|)}
\,,\ee
where $w$ is called the kernel. For $w(z)=\Theta(\epsilon-z)$ where $\Theta$ is
the Heaviside step function, we return to Eq.(\ref{eq:predict.lazy}).

\begin{figure}
\centerline{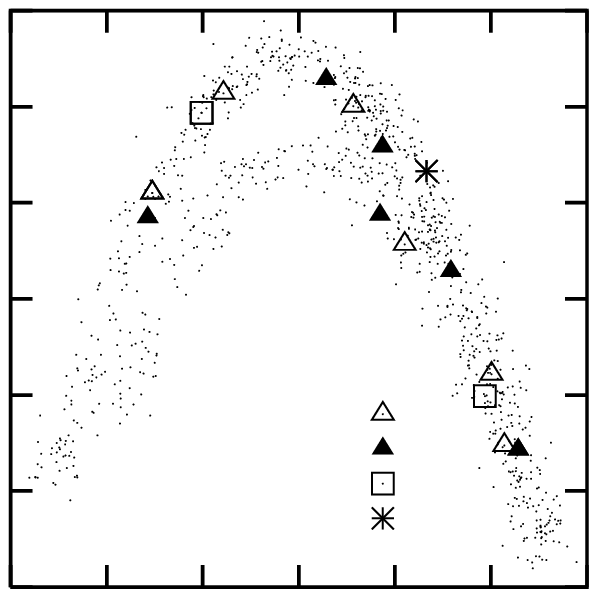}
\caption[]{\label{fig:upo}\small
   Orbits of period six, or a sub-period thereof, of the H\'enon map,
   determined from noisy data. The H\'enon attractor does not have a period
   three orbit.}
\end{figure}

\subsection{Finding unstable periodic orbits}\label{sec:upo}
As an application of simple nonlinear phase space prediction, let us discuss a
method to locate unstable periodic orbits embedded in a chaotic attractor. This
is not the place to review the existing methods to solve this problem, some
references include~\cite{auerbach,biham,so,schmelcher}. The {\tisean} package
contains a routine that implements the requirement that for a period $p$ orbit
$\{\tilde\ss_n,n=1,\ldots,p\}$ of a dynamical system like Eq.(\ref{eq:map}) acting on
delay vectors
\be\label{eq:upo}
   \tilde\ss_{n+1}=\ff(\tilde\ss_n), 
   \quad n=1,\ldots,p, 
   \quad \tilde\ss_{p+1}\equiv\tilde\ss_1
\,.\ee
With unit delay, the $p$ delay vectors contain $p$ different scalar entries,
and Eq.(\ref{eq:upo}) defines a root of a system of $p$ nonlinear equations in
$p$ dimensions. Multidimensional root finding is not a simple problem.  The
standard Newton method has to be augmented by special tricks in order to
converge globally. Some such tricks, in particular means to select different
solutions of Eq.(\ref{eq:upo}) are implemented in~\cite{schmelcher}.  Similarly
to the problems encountered in nonlinear noise reduction, solving
Eq.(\ref{eq:upo}) {\em exactly} is particularly problematic since $\ff(\cdot)$
is unknown and must be estimated from the data. In Ref.~\cite{so}, approximate
solutions are found by performing just one iteration of the Newton method for
each available time series point. We prefer to look for a {\em least squares}
solution by minimizing
\be\label{eq:upo2}
   \sum_{n=1}^p \|\tilde\ss_{n+1}-\ff(\tilde\ss_n)\|^2,
   \quad \tilde\ss_{p+1}\equiv\tilde\ss_1
\ee
instead. The routine {\UPO} uses a standard Levenberg-Marquardt algorithm to
minimize (\ref{eq:upo2}). For this it is necessary that $\ff(\cdot)$ is
smooth. Therefore we cannot use the simple nonlinear predictor based on locally
constant approximations and we have to use a smooth kernel version,
Eq.(\ref{eq:predict.kernel}), instead. With $w(z)=exp(-z^2/2h^2)$, the kernel
bandwidth $h$ determines the degree of smoothness of $\ff(\cdot)$. Trying to
start the minimization with all available time series segments will produce a
number of false minima, depending on the value of $h$. These have to be
distinguished from the true solutions by inspection. On the other hand, we can
reach solutions of Eq.(\ref{eq:upo}) which are not closely visited in the time
series at all, an important advantage over close return
methods~\cite{auerbach}.

It should be noted that, depending on $h$, we may always find good minima of
(\ref{eq:predict.kernel}), even if no solution of Eq.(\ref{eq:upo}), or not
even a truly deterministic dynamics exists. Thus the finding of unstable
periodic orbits in itself is not a strong indicator of determinism.
We may however use the cycle locations or stabilities as a discriminating
statistics in a test for nonlinearity, see Sec.~\ref{sec:surro}.
While the orbits themselves are found quite easily, it is surprisingly
difficult to obtain reliable estimates of their stability in the presence of
noise. In {\UPO}, a small perturbation is iterated along the orbit and the
unstable eigenvalue is determined by the rate of its separation from the
periodic orbit. 

The user of {\UPO} has to specify the embedding dimension, the period (which
may also be smaller) and the kernel bandwidth. For efficiency, one may choose
to skip trials with very similar points. Orbits are counted as distinct only
when they differ by a specified amount. The routine finds the orbits, their
expanding eigenvalue, and possible sub-periods. Figure~\ref{fig:upo} shows the
determination of all period six orbits from 1000 iterates of the H\'enon map,
contaminated by 10\% Gaussian white noise.

\subsection{Locally linear prediction}
If there is a good reason to assume that the relation $s_{n+1}=f(\ss_n)$ is
fulfilled by the experimental data in good approximation (say, within 5\%) for
some unknown $f$ and that $f$ is smooth, predictions can be improved by
fitting local linear models. They can be considered as the local Taylor
expansion of the unknown $f$, and are easily determined by minimizing
\be\label{eq:predict.loclin1}
   \sigma^2=\sum_{\ss_j\in{\cal U}_n} (s_{j+1}-\aa_n\ss_j - b_n)^2
\ee
with respect to $\aa_n$ and $b_n$, where ${\cal U}_n$ is the
$\epsilon$-neighborhood of $\ss_n$, excluding $\ss_n$ itself, as before. Then,
the prediction is $\hat s_{n+1}=\aa_n\ss_n +b_n$. The minimization problem can
be solved through a set of coupled linear equations, a standard linear algebra
problem.  This scheme is implemented in {\ONESTEP}. For moderate noise levels
and time series lengths this can give a reasonable improvement over {\ZEROTH}
and {\PREDICT}. Moreover, as discussed in Sec.\ref{sec:lyap}, these linear maps
are needed for the computation of the Lyapunov spectrum. Locally linear
approximation was introduced in~\cite{Eckmann,fsid0}. We should note that the
straight least squares solution of Eq.(\ref{eq:predict.loclin1}) is not always
optimal and a number of strategies are available to regularize the problem if 
the matrix becomes nearly singular and to remove the bias due to the errors in
the ``independent'' variables. These strategies have in common that any
possible improvement is bought with considerable complication of the procedure,
requiring subtle parameter adjustments. We refer the reader to
Refs.~\cite{kugiLL,jaeger} for advanced material.

\begin{figure}[t]
\centerline{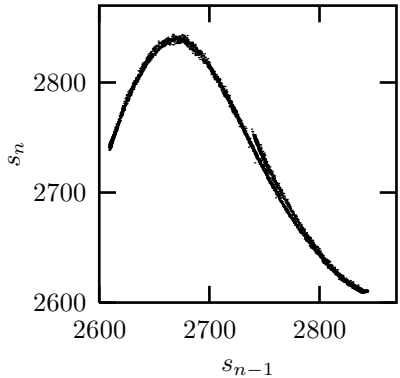}
\caption[]{\label{fig:predict.INOnstep}\small 
   Time delay representation of 5000 iterations of the local linear predictor
   {\NSTEP} in two dimensions, starting from the last delay vector of
   Fig.~\ref{fig:delay.Poincare}.}
\end{figure}

In Fig.~\ref{fig:predict.INOnstep} we show iterated predictions of the
Poincar\'e map data from the CO$_2$ laser (Fig.~\ref{fig:delay.Poincare}) in a
delay representation (using {\NSTEP} in two dimensions). The resulting data do
not only have the correct marginal distribution and power spectrum, but also
form a perfect skeleton of the original noisy attractor. There are
of course artefacts due to noise and the roughness of this approach,
but there are good reasons to assume that the line-like substructure
reflects fractality of the unperturbed system.

\begin{figure}[t]
\centerline{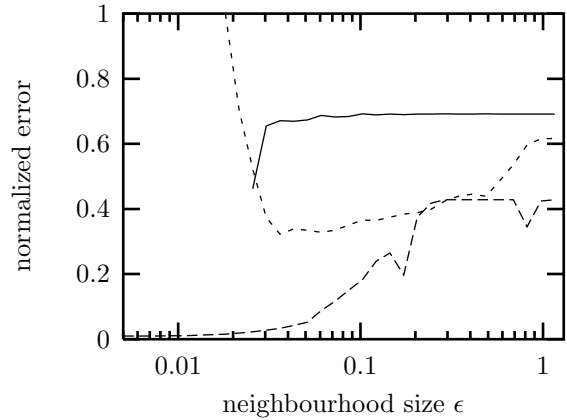}
\caption[]{\label{fig:predict.casdagli}\small 
   The Casdagli test for nonlinearity: The rms prediction error of local linear
   models as a function of the neighborhood size $\epsilon$. Lower curve: The
   CO$_2$ laser data. These data are obviously highly deterministic in
   $m$=4 dimensions and with lag $\tau$=6.  Central curve: The breath rate data
   shown in Fig.~\ref{fig:b} with $m$=4 and $\tau$=1. Determinism is weaker
   (presumably due to a much higher noise level), but still the nonlinear
   structure is dominant. Upper curve: Numerically generated data of an AR(5)
   process, a linearly correlated random process ($m$=5, $\tau$=1).}
\end{figure}

Casdagli~\cite{Casdagli_royal} suggested to use local linear models as a test 
for nonlinearity: He computed the average forecast error as a
function of the neighborhood size on which the fit for $\aa_n$ and $b_n$ is
performed. If the optimum occurs at large neighborhood sizes, the data are (in
this embedding space) best described by a linear stochastic process, whereas an
optimum at rather small sizes supports the idea of the existence of a nonlinear
almost deterministic equation of motion. This protocol is implemented in the
routine {\LLAR}, see Fig.~\ref{fig:predict.casdagli}.

\subsection{Global function fits}
The local linear fits are very flexible, but can go wrong on parts of the
phase space where the points do not span the available space dimensions and
where the inverse of the matrix involved in the solution of the minimization
does not exist.  Moreover, very often a large set of different linear maps is
unsatisfying. Therefore many authors suggested to fit global nonlinear
functions to the data, i.e.\ to solve
\be\label{eq:predict.global}
   \sigma^2=\sum_n (s_{n+1}-f_p(\ss_n))^2
\,,\ee
where $f_p$ is now a nonlinear function in closed form with parameters $p$,
with respect to which the minimization is done. Polynomials, radial basis
functions, neural nets, orthogonal polynomials, and many other approaches have
been used for this purpose. The results depend on how far the chosen ansatz
$f_p$ is suited to model the unknown nonlinear function, and on how well the
data are deterministic at all. We included the routines {\RBF} and {\POLYNOM}
in the {\tisean} package, where $f_p$ is modeled by radial basis
functions~\cite{rbf,lenny_rbf} and polynomials~\cite{Casdagli_pred},
respectively. The advantage of these two models is that the parameters $p$
occur linearly in the function $f$ and can thus be determined by simple linear algebra, and the solution is unique. Both features are lost for
models where the parameters enter nonlinearly.  

\begin{figure}
\centerline{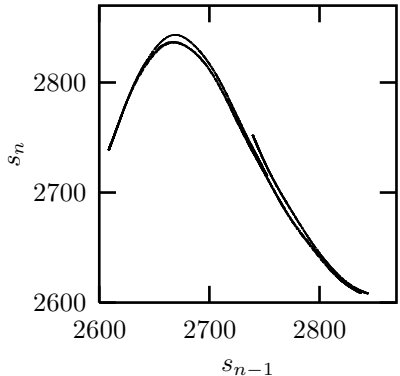}
\caption[]{\label{fig:predict.INO.rbf}\small 
   Attractor obtained by iterating the model that has been obtained by a fit 
   with 40 radial basis functions in two dimensions to the time series
   shown in Fig.~\ref{fig:delay.Poincare}. Compare also 
   Fig.~\ref{fig:predict.INOnstep}.}
\end{figure}

In order to make global nonlinear predictions, one has to supply the embedding
dimension and time delay as usual. Further, for {\POLYNOM} the order of the
polynomial has to be given. The program returns the coefficients of the model.
In {\RBF} one has to specify the number of basis functions to be distributed
on the data.  The width of the radial basis functions (Lorentzians in our
program) is another parameter, but since the minimization is so fast, the
program runs many trial values and returns parameters for the
best. Figure~\ref{fig:predict.INO.rbf} shows the result of a fit to the CO$_2$
laser time series (Fig.~\ref{fig:delay.Poincare}) with radial basis functions.

If global models are desired in order to infer the structure and properties of
the underlying system, they should be tested by iterating them. The prediction
errors, although small in size, could be systematic and thus repel the
iterated trajectory from the range where the original data are located.  It
can be useful to study a dependence of the size or the sign of the prediction
errors on the position in the embedding space, since systematic errors can be
reduced by a different model.  Global models are attractive because they yield
closed expressions for the full dynamics. One must not forget, however, that
these models describe the observed process only in regions of the space which
have been visited by the data. Outside this area, the shape of the model
depends exclusively on the chosen ansatz. In particular, polynomials diverge
outside the range of the data and hence can be unstable under iteration.

\section{Nonlinear noise reduction}\label{sec:noise}
Filtering of signals from nonlinear systems requires the use of special
methods since the usual spectral or other linear filters may interact
unfavorably with the nonlinear structure. Irregular signals from nonlinear
sources exhibit genuine broad band spectra and there is no justification to
identify any continuous component in the spectrum as noise. Nonlinear noise
reduction does not rely on frequency information in order to define the
distinction between signal and noise. Instead, structure in the reconstructed
phase space will be exploited. General serial dependencies among the
measurements $\{s_n\}$ will cause the delay vectors $\{\Ss_n\}$ to fill the
available $m$-dimensional embedding space in an inhomogeneous way. Linearly
correlated Gaussian random variables will for example be distributed according
to an anisotropic multivariate Gaussian distribution. Linear geometric
filtering in phase space seeks to identify the principal directions of this
distribution and project onto them, see Sec.~\ref{sec:svdfilter}. Nonlinear
noise reduction takes into account that nonlinear signals will form curved
structures in delay space.  In particular, noisy {\em deterministic} signals
form smeared-out lower dimensional manifolds. Nonlinear phase space filtering
seeks to identify such structures and project onto them in order to reduce
noise.

There is a rich literature on nonlinear noise reduction methods. Two articles
of review character are available, one by Kostelich and Schreiber~\cite{ks},
and one by Davies~\cite{Davies}. We refer the reader to these articles for
further references and for the discussion of approaches not described in the
present article. Here we want to concentrate on two approaches that represent
the geometric structure in phase space by local approximation. The first and
simplest does so to constant order, the more sophisticated uses local linear
subspaces plus curvature corrections.

\subsection{Simple nonlinear noise reduction}
The simplest nonlinear noise reduction algorithm we know of replaces
the central coordinate of each embedding vector by the local average of this
coordinate. This amounts to a locally constant approximation of the dynamics
and is based on the assumption that the dynamics is continuous. The algorithm
is described in~\cite{lazy}, a similar approach is proposed in~\cite{Arkady}.
In an unstable, for example chaotic, system, it is essential not to replace 
the first and last coordinates of the embedding vectors by local averages.
Due to the instability, initial errors in these coordinates are magnified
instead of being averaged out. 

This noise reduction scheme is implemented quite easily. First an embedding
has to be chosen. Except for extremely oversampled data, it is advantageous to
choose a short time delay. The program {\LAZY} always uses unit delay.  The
embedding dimension $m$ should be chosen somewhat higher than that required by
the embedding theorems. Then for each embedding vector $\{\Ss_n\}$, a
neighborhood ${\cal U}_{\epsilon}^{(n)}$ is formed in phase space containing
all points $\{\Ss_{n'}\}$ such that $\|\Ss_n-\Ss_{n'}\| < \epsilon$. The
radius of the neighborhoods $\epsilon$ should be taken large enough in order
to cover the noise extent, but still smaller than a typical curvature radius.
These conditions cannot always be fulfilled simultaneously, in which case one
has to repeat the process with several choices and carefully evaluate the
results. If the noise level is substantially smaller than the typical radius
of curvature, neighborhoods of radius about 2-3 times the noise level gave the
best results with artificial data. For each embedding vector
$\Ss_n=(s_{n-(m-1)}, \ldots, s_n)$ (the delay time has been set to unity), a
corrected middle coordinate $\hat{s}_{n-m/2}$ is computed by averaging over
the neighborhood ${\cal U}_{\epsilon}^{(n)}$:
\be\label{eq:lazy}
   \hat{s}_{n-m/2} = {1 \over |{\cal U}_{\epsilon}^{(n)}|} 
      \sum_{\Ss_{n'}\in {\cal U}_{\epsilon}^{(n)}} s_{n'-m/2}
\,.\ee
After one complete sweep through the time series, all measurements $s_n$ are
replaced by the corrected values $\hat{s}_n$. Of course, for the first and last
$(m-1)/2$ points (if $m$ is odd), no correction is available. The average
correction can be taken as a new neighborhood radius for the next iteration.
Note that the neighborhood of each point at least contains the point itself.
If that is the only member, the average Eq.(\ref{eq:lazy}) is simply the
uncorrected measurement and no change is made. Thus one can safely perform
multiple iterations with decreasing values of $\epsilon$ until no further
change is made.

\begin{figure}[t]
\begin{center}
   ~\\[-15pt]
   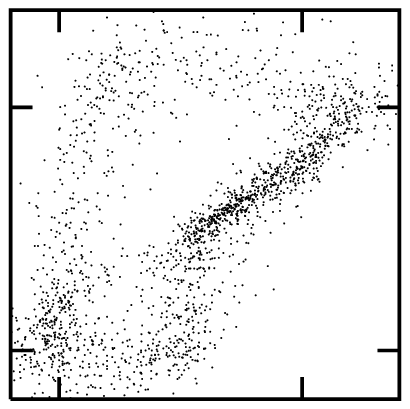
   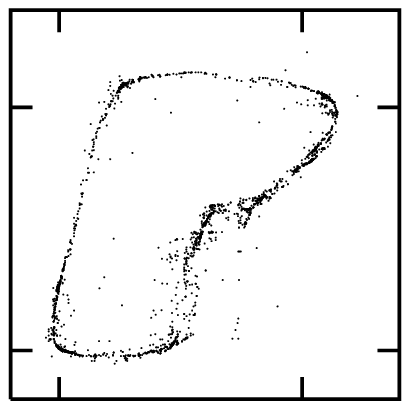
\end{center}
\caption[]{\label{fig:b}\small
  Simple nonlinear noise reduction of human breath rate data. Three iterations
  have been carried out, staring with neighborhoods of size 0.4~units.
  Embeddings in 7 dimensions at unit delay have been used. Arguably, the
  resulting series (lower panel) is less noisy. However, in
  Sec.~\ref{sec:surro} we will show evidence that the noise is not just
  additive and independent of the signal. }
\end{figure}

Let us illustrate the use of this scheme with an example, a recording of the
air flow through the nose of a human as an indicator of breath
activity. (The
data is part of data set B of the Santa Fe time series contest held in
1991/92~\cite{SFI}, see Rigney et al.~\cite{b.dat} for a description.)
The result of simple nonlinear noise reduction is shown in Fig.~\ref{fig:b}.

\subsection{Locally projective nonlinear noise reduction}
A more sophisticated method makes use of the hypotheses that the measured data
is composed of the output of a low-dimensional dynamical system and of random
or high-dimensional noise. This means that in an arbitrarily high-dimensional
embedding space the deterministic part of the data would lie on a
low-dimensional manifold, while the effect of the noise is to spread the data
off this manifold. If we suppose that the amplitude of the noise is
sufficiently small, we can expect to find the data distributed closely around
this manifold. The idea of the projective nonlinear noise reduction scheme is
to identify the manifold and to project the data onto it. The strategies
described here go back to Ref.~\cite{on}. A realistic case study is detailed
in Ref.~\cite{buzug}.

Suppose the dynamical system, Eq.~(\ref{eq:ode}) or Eq.~(\ref{eq:map}), form a
$q$-dimensional manifold $\cal M$ containing the trajectory. According to the
embedding theorems, there exists a one-to-one image of the attractor
 in the embedding space, if the embedding dimension is sufficiently
high. Thus, if the measured time series were not corrupted with noise, all the
embedding vectors ${\bf s}_n$ would lie inside another manifold
$\tilde{\cal M}$ in the embedding space. Due to the noise
this condition is no longer fulfilled. The idea of the locally projective noise
reduction scheme is that for each ${\bf s}_n$ there exists a correction
${\bf\Theta}_n$, with $\|{\bf\Theta}_n\|$ small, in such a way that ${\bf
s}_n-{\bf \Theta}_n\in\tilde{\cal M}$ and that ${\bf \Theta}_n$ is orthogonal
on $\tilde{\cal M}$. Of course a projection to the manifold can only be a
reasonable concept if the vectors are embedded in spaces which are higher
dimensional than the manifold $\tilde{\cal M}$. Thus we have to over-embed in
$m$-dimensional spaces with $m>q$.

The notion of orthogonality depends on the metric used. Intuitively one would
think of using the Euclidean metric. But this is not necessarily the best
choice. The reason is that we are working with delay vectors which contain
temporal information.  Thus even if the middle parts of two delay
vectors are close, the late parts could be far away from each other due to the
influence of the positive Lyapunov exponents, while the first parts could
diverge due the negative ones. Hence it is usually desirable to correct only
the center part of delay vectors and leave the outer parts mostly unchanged,
since their divergence is not only a consequence of the noise, but also of the
dynamics itself. It turns out that for most applications it is sufficient to
fix just the first and the last component of the delay vectors and correct the
rest. This can be expressed in terms of a metric tensor ${\bf P}$ which we
define to be~\cite{on}
\be
   {\bf P}_{ij}=\left\{\begin{array}{l@{\quad:\quad}l}
   1 & i=j\quad\mbox{and}\quad 1<i,j<m\\
   0 & \mbox{elsewhere}
\end{array}\right.
\,,\ee
where $m$ is the dimension of the ``over-embedded'' delay vectors.

Thus we have to solve the minimization problem
\be
   \sum_i \left({\bf\Theta}_i {\bf P^{-1}\Theta}_i\right)
   \stackrel{!}{=}\mbox{min}
\ee
with the constraints
\be
   {\bf a}_n^i\left({\bf s}_n-{\bf\Theta}_n\right)
   +b^i_n=0\qquad\mbox{for}\quad i=q+1,\ldots,m
\ee
and
\be
   {\bf a}_n^i{\bf P}{\bf a}_n^j=\delta_{ij}
\ee
where the ${\bf a}_n^i$ are the normal vectors of $\tilde{\cal M}$ at the point
${\bf s}_n-{\bf \Theta}_n$.

This ideas are realized in the programs {\GHKSS}, {\PROJECT}, and {\NOISE} in
{\tisean}. While the first two work as {\em a posteriori} filters on complete
data sets, the last one can be used in a data stream. This means that it is
possible to do the corrections online, while the data is coming in (for more
details see section~\ref{subsec.noise_stream}).  All three algorithms mentioned
above correct for curvature effects. This is done by either post-processing the
corrections for the delay vectors ({\GHKSS}) or by preprocessing the centres of
mass of the local neighborhoods ({\PROJECT}).

The idea used in the {\GHKSS} program is the following. Suppose the manifold
were strictly linear. Then, provided the noise is white, the corrections in the
vicinity of a point on the manifold would point in all directions with the same
probability. Thus, if we added all the corrections ${\bf \Theta}$ we expect
them to sum to zero (or $\langle{\bf\Theta}\rangle={\bf O}$). On the other
hand, if the manifold is curved, we expect that there is a trend towards the
centre of curvature ($\langle{\bf\Theta}\rangle =
{\bf\Theta}_{\mbox{\scriptsize av}}$). Thus, to correct for this trend each
correction ${\bf\Theta}$ is replaced by
${\bf\Theta}-{\bf\Theta}_{\mbox{\scriptsize av}}$.

A different strategy is used in the program {\PROJECT}. The projections are
done in a local coordinate system which is defined by the condition that the
average of the vectors in the neighborhood is zero. Or, in other words, the
origin of the coordinate systems is the centre of mass $\langle{\bf
s}_n\rangle_{\cal U}$ of the neighborhood $\cal U$. This centre of mass has a
bias towards the centre of the curvature~\cite{KantzSchreiber}. Hence, a
projection would not lie on the tangent at the manifold, but on a secant. Now
we can compute the centre of mass of these points in the neighborhood of ${\bf
s}_n$. Let us call it $\langle\langle{\bf s}_n\rangle\rangle_{\cal U}$. Under
fairly mild assumptions this point has twice the distance from the manifold
then $\langle{\bf s}_n\rangle_{\cal U}$. To correct for the bias the origin of
the local coordinate system is set to the point: $\langle\langle{\bf
s}_n\rangle\rangle_{\cal U} - 2\langle{\bf s}_n\rangle_{\cal U}$.

The implementation and use of locally projective noise reduction as realized
in {\PROJECT} and {\GHKSS} is described in detail in Refs.~\cite{on,buzug}.
Let us recall here the most important parameters that have to be set
individually for each time series. The embedding parameters are usually chosen
quite differently from other applications since considerable over-embedding may
lead to better noise averaging. Thus, the delay time is preferably set to unity
and the embedding dimension is chosen to provide embedding windows of
reasonable lengths. Only for highly oversampled data (like the
magneto-cardiogram, Fig.~\ref{fig:mcgnoise}, at about 1000 samples per cycle),
larger delays are necessary so that a substantial fraction of a cycle can be
covered without the need to work in prohibitively high dimensional spaces.
Next, one has to decide how many dimensions $q$ to leave for the manifold
supposedly containing the attractor. The answer partly depends on the purpose
of the experiment. Rather brisk projections can be optimal in the sense of
lowest residual deviation from the true signal. Low rms error can however 
coexist with systematic distortions of the attractor structure. Thus for a
subsequent dimension calculation, a more conservative choice would be in order.
Remember however that points are only moved {\em towards} the local linear
subspace and too low a value of $q$ does not do as much harm as may be though.

\begin{figure}[t]
\begin{center}
   ~\\[-10pt]
   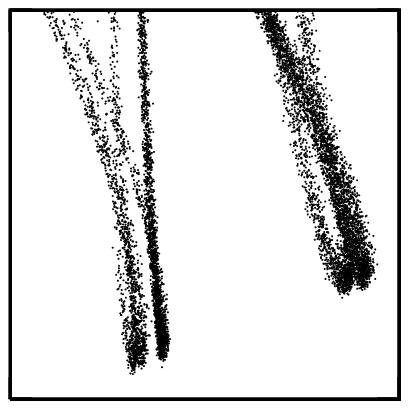\\
   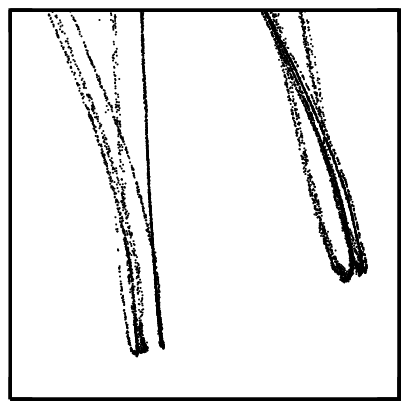
\end{center}
\caption[]{\label{fig.noise_opt_raser}\small 
   Two-dimensional representation of the NMR Laser data  (top) and the 
   result of the {\GHKSS} algorithm (bottom) after three iterations.}
\end{figure}

The noise amplitude to be removed can be selected to some degree by the choice
of the neighborhood size. In fact, nonlinear projective filtering can be seen
independently of the dynamical systems background as filtering by amplitude
rather than by frequency or shape. To allow for a clear separation of noise and
signal directions locally, neighborhoods should be at least as large as the
supposed noise level, rather larger. This of course competes with curvature
effects. For small initial noise levels, it is recommended to also specify a
minimal number of neighbors in order to permit stable linearizations.
Finally, we should remark that in successful cases most of the filtering is
done within the first one to three iterations. Going further is potentially
dangerous since further corrections may lead mainly to distortion.
One should watch the rms correction in each iteration and stop as soon as it
doesn't decrease substantially any more.

As an example for nonlinear noise reduction we treat the data obtained from an
NMR laser experiment~\cite{raser}. Enlargements of two-dimensional delay
representations of the data are shown in Fig.~\ref{fig.noise_opt_raser}. The
upper panel shows the raw experimental data which contains about 1.1\% of
noise. The lower panel was produced by applying three iterations of the noise
reduction scheme. The embedding dimension was $m=7$, the vectors were projected
down to two dimensions. The size of the local neighborhoods were chosen such
that at least 50 neighbors were found.  One clearly sees that the fractal
structure of the attractor is resolved fairly well.

\begin{figure}[t]
\begin{center}
   ~\\[-10pt]
   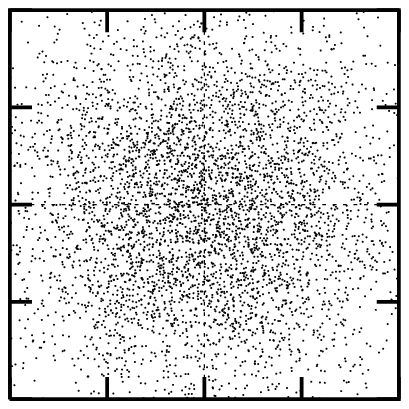\\
   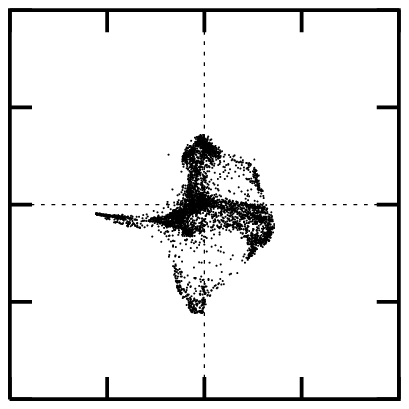
\end{center}
\caption[]{\label{fig.noise_opt_breath}\small 
   Two-dimensional representation of a pure Gaussian process (top) and the
   outcome of the {\GHKSS} algorithm (bottom) after 10 iterations. Projections
   from $m=7$ down to two dimensions were performed.}
\end{figure}

The main assumption for this algorithm to work is that the data is well
approximated by a low-dimensional manifold. If this is not the case it is
unpredictable what results are created by the algorithm. In the absence of a
real manifold, the algorithm must pick statistical fluctuations and spuriously
interprets them as structure.  Figure~\ref{fig.noise_opt_breath} shows a result
of the {\GHKSS} program for pure Gaussian noise. The upper panel shows a delay
representation of the original data, the lower shows the outcome of applying
the algorithm for 10 iterations. The structure created is purely artifical and
has nothing to do with structures in the original data. This means that if one
wants to apply one of the algorithms, one has to carefully study the results.
If the assumptions underlying the algorithms are not fulfilled in principle
anything can happen. One should note however, that the performance of the
program itself indicates such spurious behavior. For data which is indeed well
approximated by a lower dimensional manifold, the average corrections applied
should rapidly decrease with each successful iteration. This was the case with
the NMR laser data and in fact, the correction was so small after three
iteration that we stopped the procedure. For the white noise data, the
correction only decreased at a rate that corresponds to a general shrinking of
the point set, indicating a lack of convergence towards a genuine low
dimensional manifold. Below, we will give an example where an approximating
manifold is present without pure determinism. In that case, projecting onto the
manifold does reduce noise in a reasonable way. See Ref.~\cite{danger} for
material on the dangers of geometric filtering. 

\subsection{Nonlinear noise reduction in a data stream}
\label{subsec.noise_stream}
\begin{figure}[t]
\begin{center}
   ~\\[-10pt]
   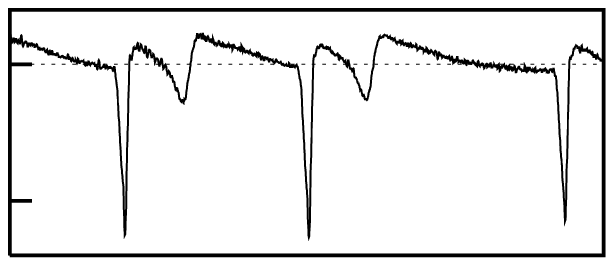\\[-17pt]
   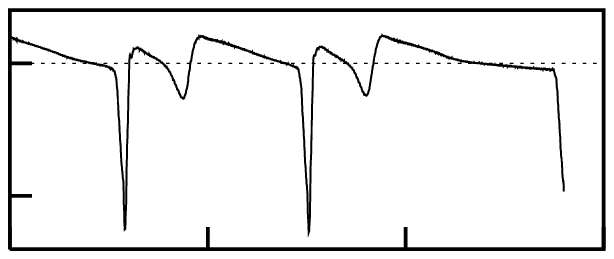
\end{center}
\caption[]{\label{fig:mcgnoise}\small
   Real time nonlinear projective filtering of a magneto-cardiogram time
   series. The top panel shows the unfiltered data. Bottom: Two iterations were
   done using projections from $m=10$ down to $q=2$ dimensions (delay
   0.01~s). Neighborhoods were limited to a radius of 0.1 units (0.05 in the
   second iteration) and to maximally 200 points. Neighbors were only sought
   up to 5~s back in time. Thus the first 5~s of data are not filtered
   optimally and are not shown here. Since the output of each iteration leaps
   behind its input by one delay window the last 0.2~s cannot be processed
   given the data in the upper panel.}
\end{figure}

In Ref.~\cite{Filter}, a number of modifications of the above procedure have
been discussed which enable the use of nonlinear projective filtering in a data
stream. In this case, only points in the past are available for the formation
of neighborhoods. Therefore the neighbor search strategy has to be modified.
Since the algorithm is described in detail in Ref.~\cite{Filter}, we only give
an example of its use here. Figure~\ref{fig:mcgnoise} shows the result of
nonlinear noise reduction on a magneto-cardiogram (see Figs.~\ref{fig:mcgd} 
and~\ref{fig:mcg_pc}) with the program {\NOISE}. The same program has also been
used successfully for the extraction of the fetal ECG~\cite{fetal2}.

\section{Lyapunov exponents}\label{sec:lyap}
Chaos arises from the exponential growth of infinitesimal perturbations,
together with global folding mechanisms to guarantee boundedness of the
solutions. This exponential instability is characterized by the spectrum of
Lyapunov exponents~\cite{EckRuelle}. If one assumes a local decomposition of
the phase space into directions with different stretching or contraction rates,
then the spectrum of exponents is the proper average of these local rates over
the whole invariant set, and thus consists of as many exponents as there are
space directions. The most prominent problem in time series analysis is that
the physical phase space is unknown, and that instead the spectrum is computed
in some embedding space. Thus the number of exponents depends on the
reconstruction, and might be larger than in the physical phase space. Such
additional exponents are called {\sl spurious}, and there are several
suggestions to either avoid them~\cite{spurious} or to identify them. Moreover,
it is plausible that only as many exponents can be determined from a time
series as are entering the Kaplan Yorke formula (see below). To give a simple
example: Consider motion of a high-dimensional system on a stable limit
cycle. The data cannot contain any information about the stability of this
orbit against perturbations, as long as they are exactly on the limit
cycle. For transients, the situation can be different, but then data are not
distributed according to an invariant measure and the numerical values are thus
difficult to interpret. Apart from these difficulties, there is one relevant
positive feature: Lyapunov exponents are invariant under smooth transformations
and are thus independent of the measurement function or the embedding
procedure. They carry a dimension of an inverse time and have to be normalized
to the sampling interval.

\begin{figure}[t]
\centerline{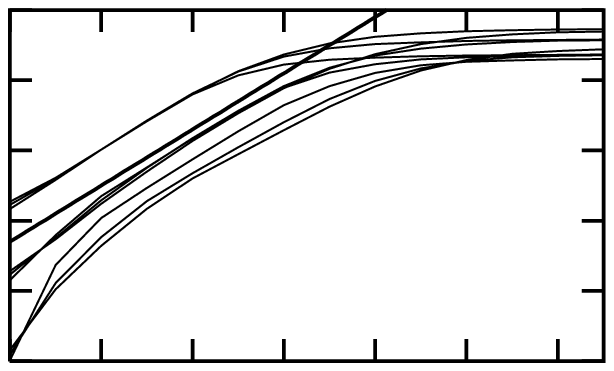}
\centerline{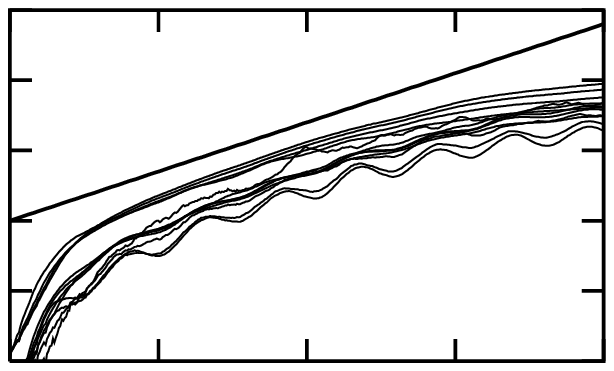}
\caption[]{\label{fig:lyap.1}\small 
  Estimating the maximal Lyapunov exponent of the CO$_2$ laser data. The
  top panel shows results for the Poincar\'e map data, where the average time
  interval $T_{\mbox{\scriptsize av}}$ is 52.2 samples of the flow, and the
  straight line indicates $\lambda=0.38$. For comparison: The iteration of the
  radial basis function model of Fig.~\ref{fig:predict.INO.rbf} yields
  $\lambda$=0.35. Bottom panel: Lyapunov exponents determined directly from the
  flow data. The straight line has slope
  $\lambda=0.007$. In good approximation, $\lambda_{\mbox{\scriptsize
  map}}=\lambda_{\mbox{\scriptsize flow}}T_{\mbox{\scriptsize av}}$. Here, the
  time window $w$ to suppress correlated neighbors has been set to 1000, and
  the delay time was 6~units.}
\end{figure}

\subsection{The maximal exponent}\label{sec:lyapmax}
The maximal Lyapunov exponent can be determined without the explicit
construction of a model for the time series. A reliable characterization
requires that the independence of embedding parameters and the exponential law
for the growth of distances are checked~\cite{Holger,rose} explicitly.
Consider the representation of the time series data as a trajectory in the
embedding space, and assume that you observe a very close return $\ss_{n'}$ to
a previously visited point $\ss_n$.  Then one can consider the distance
$\Delta_0= \ss_n-\ss_{n'}$ as a small perturbation, which should grow
exponentially in time. Its future can be read from the time series:
$\Delta_l=\ss_{n+l}-\ss_{n'+l}$. If one finds that $|\Delta_l|\approx
\Delta_0 e^{\lambda l}$ then $\lambda$ is (with probability one) the maximal
Lyapunov exponent. In practice, there will be fluctuations because of many
effects, which are discussed in detail in~\cite{Holger}. Based on this
understanding, one can derive a robust consistent and unbiased estimator for
the maximal Lyapunov exponent.  One computes
\be\label{eq:S}
   S(\epsilon,m,t)=\left\langle \ln\left(\frac{1}{|{\cal
   U}_n|}\sum_{\ss_{n'}\in{\cal U}_n} |s_{n+t}-s_{n'+t}|\right)\right\rangle_n
\,.\ee
If $S(\epsilon,m,t)$ exhibits a linear increase with identical slope for all
$m$ larger than some $m_0$ and for a reasonable range of $\epsilon$, then this
slope can be taken as an estimate of the maximal exponent $\lambda_1$. 

The formula is implemented in the routines {\LYAPK} and {\LYAPUNOV} in a
straightforward way. (The program {\LYAPR} implements the very similar
algorithm of Ref.~\cite{rose} where only the closest neighbor is followed for
each reference point. Also, the Euclidean norm is used.)  Apart from parameters
characterizing the embedding, the initial neighborhood size $\epsilon$ is of
relevance: The smaller $\epsilon$, the large the linear range of $S$, if there
is one. Obviously, noise and the finite number of data points limit $\epsilon$
from below. The default values of {\LYAPK} are rather reasonable for map-like
data. It is not always necessary to extend the average in Eq.(\ref{eq:S}) over
the whole available data, reasonable averages can be obtained already with a
few hundred reference points $\Ss_n$. If some of the reference points have very
few neighbors, the corresponding inner sum in Eq.(\ref{eq:S}) is dominated by
fluctuations. Therefore one may choose to exclude those reference points which
have less than, say, ten neighbors. However, discretion has to be applied with
this parameter since it may introduce a bias against sparsely populated
regions. This could in theory affect the estimated exponents due to
multifractality. Like other quantities, Lyapunov estimates may be affected by
serial correlations between reference points and neighbors. Therefore, a
minimum time for $|n-n'|$ can and should be specified here as well. See also
Sec.\ref{sec:dimension}.

\begin{figure}[t]
\centerline{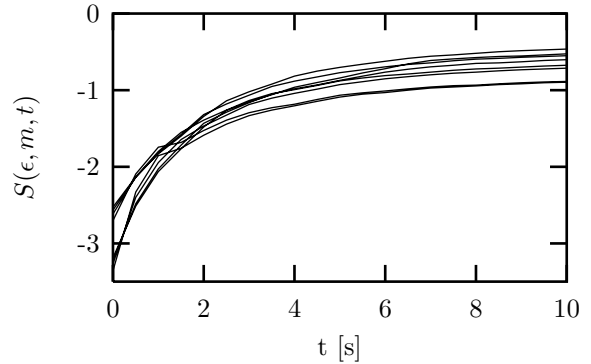}
\caption[]{\label{fig:lyap.2}\small 
   The breath rate data (c.f. Fig.~\ref{fig:b}) exhibit no linear increase,
   reflecting the lack of exponential divergence of nearby trajectories.}
\end{figure}

Let us discuss a few typical outcomes. The data underlying the top panel of
Fig.~\ref{fig:lyap.1} are the values of the maxima of the CO$_2$ laser
data. Since this laser exhibits low dimensional chaos with a reasonable noise
level, we observe a clear linear increase in this semi-logarithmic plot,
reflecting the exponential divergence of nearby trajectories. The exponent is
$\lambda\approx0.38$ per iteration (map data!), or, when introducing the
average time interval, 0.007 per $\mu$s. In the bottom panel we show the result
for the same system, but now computed on the original flow-like data with a
sampling rate of 1~MHz. As additional structure, an initial steep increase and
regular oscillations are visible. The initial increase is due to non-normality
and effects of alignment of distances towards the locally most unstable
direction, and the oscillations are an effect of the locally different
velocities and thus different densities. Both effects can be much more dramatic
in less favorable cases, but as long as the regular oscillations possess a
linearly increasing average, this can be taken as the estimate of the Lyapunov
exponent. Normalizing by the sampling rate, we again find $\lambda\approx$
0.007 per $\mu$s, but it is obvious that the linearity is less pronounced then
for the map-like data.  Finally, we show in Fig.~\ref{fig:lyap.2} an example
of a negative result: We study the human breath rate data used before. No
linear part exists, and one cannot draw any reasonable conclusion. It is worth
considering the figure on a doubly logarithmic scale in order to detect a power
law behavior, which, with power 1/2, could be present for a diffusive growth
of distances. In this particular example, there is no convincing power law
either. 

\subsection{The Lyapunov spectrum}
The computation of the full Lyapunov spectrum requires considerably more effort
than just the maximal exponent. An essential ingredient is some estimate of the
local Jacobians, i.e.\ of the linearized dynamics, which rules the growth of
infinitesimal perturbations. One either finds it from direct fits of local
linear models of the type $s_{n+1}=\aa_n\ss_n + b_n$, such that the first row
of the Jacobian is the vector $\aa_n$, and $(\JJ)_{ij}=\delta_{i-1,j}$ for
$i=2,\ldots,m$, where $m$ is the embedding dimension. The $\aa_n$ is given by
the least squares minimization $\sigma^2=\sum_l (s_{l+1} - \aa_n\ss_l -b_n)^2$
where $\{\ss_l\}$ is the set of neighbors of $\ss_n$~\cite{Eckmann,sano}.  Or
one constructs a global nonlinear model and computes its local Jacobians by
taking derivatives. In both cases, one multiplies the Jacobians one by one,
following the trajectory, to as many different vectors $\uu_k$ in tangent space
as one wants to compute Lyapunov exponents. Every few steps, one applies a
Gram-Schmidt orthonormalization procedure to the set of $\uu_k$, and
accumulates the logarithms of their rescaling factors. Their average, in the
order of the Gram-Schmidt procedure, give the Lyapunov exponents in descending
order. The routine {\LYAPSPEC} uses this method, which goes back to~\cite{sano}
and~\cite{Eckmann}, employing local linear fits. Apart from the problem of
spurious exponents, this method contains some other pitfalls: It {\sl assumes}
that there exist well defined Jacobians, and does not test for their
relevance. In particular, when attractors are thin in the embedding space, some
(or all) of the local Jacobians might be estimated very badly. Then the whole
product can suffer from these bad estimates and the exponents are
correspondingly wrong. Thus the global nonlinear approach can be superior, if a
modeling has been successful, see Sec.~\ref{sec:predict}.

In Table~\ref{tab:Lyap1} we show the exponents of the stroboscopic NMR
laser data in a three dimensional embedding as a function of the neighborhood
size.  Using global nonlinear models, we find the numbers given in the last
two rows. More material is discussed in~\cite{KantzSchreiber}. The spread of
values in the table for this rather clean data set reflects the difficulty of
estimating Lyapunov spectra from time series, which has to be done with great
care. In particular, when the algorithm is blindly applied to data from a
random process, it cannot internally check for the consistency of the
assumption of an underlying dynamical system. Therefore a Lyapunov spectrum is
computed which now is completely meaningless.

\begin{table}
\begin{center}
~\\[-5pt]
\begin{tabular}{l@{}rccc}
method & & $\lambda_1$ & $\lambda_2$ & $\lambda_3$ \\
\hline
local linear           & $k$=20  & 0.32 & -0.40 & -1.13 \\
\multicolumn{1}{c}{``} & $k$=40  & 0.30 & -0.51 & -1.21 \\
\multicolumn{1}{c}{``} & $k$=160 & 0.28 & -0.68 & -1.31 \\
\hline
\multicolumn{2}{l}{radial basis functions} & 0.27 & -0.64 & -1.31 \\
\multicolumn{2}{l}{polynomial}             & 0.27 & -0.64 & -1.15 \\
\end{tabular}
\end{center}
\caption[]{\label{tab:Lyap1}\small
    Lyapunov exponents of the NMR laser data, determined with a
    three-dimensional embedding. The algorithms described in 
    Sec.~\ref{sec:lyapmax} give $\lambda_1=0.3\pm 0.02$ for the largest
    exponent. }
\end{table}

The computation of the first part of the Lyapunov spectrum allows for some
interesting cross-checks. It was conjectured~\cite{KaplanYorke}, and is found
to be correct in most physical situations, that the Lyapunov spectrum and the
fractal dimension of an attractor are closely related. If the expanding and
least contracting directions in space are continuously filled and only one
partial dimension is fractal, then one can ask for the dimensionality of a
(fractal) volume such that it is invariant, i.e.\ such that the sum of the
corresponding Lyapunov exponents vanishes, where the last one is weighted with
the non-integer part of the dimension:
\be\label{eq:lyap.KY}
   D_{KY}= k +\frac{\sum_{i=1}^k \lambda_i}{|\lambda_{k+1}|}
\,,\ee 
where $k$ is the maximum integer such that the sum of the $k$ largest exponents
is still non-negative. $D_{KY}$ is conjectured to coincide with the information
dimension.

The Pesin identity is valid under the same assumptions and allows to compute
the KS-entropy:
\be\label{eq:lyap.Pesin}
   h_{KS}=\sum_{i=1}^m \Theta(\lambda_i) \lambda_i
\,.\ee

\section{Dimensions and entropies}\label{sec:dimension}
Solutions of dissipative dynamical systems cannot fill a volume of the phase
space, since dissipation is synonymous with a contraction of volume elements
under the action of the equations of motion. Instead, trajectories are confined
to lower dimensional subsets which have measure zero in the phase space. These
subsets can be extremely complicated, and frequently they possess a fractal
structure, which means that they are in a nontrivial way
self-similar. Generalized dimensions are one class of quantities to
characterize this fractality. The {\em Hausdorff dimension} is, from the
mathematical point of view, the most natural concept to characterize fractal
sets~\cite{EckRuelle}, whereas the {\em information dimension} takes into
account the relative visitation frequencies and is therefore more attractive for
physical systems. Finally, for the characterization of measured data, other
similar concepts, like the {\em correlation dimension}, are more useful. One
general remark is highly relevant in order to understand the limitations of any
numerical approach: dimensions characterize a set or an invariant measure whose
support is the set, whereas any data set contains only a finite number of
points representing the set or the measure. By definition, the dimension of a
finite set of points is zero. When we determine the dimension of an attractor
numerically, we extrapolate from finite length scales, where the statistics we
apply is insensitive to the finiteness of the number of data, to the
infinitesimal scales, where the concept of dimensions is defined. This
extrapolation can fail for many reasons which will be partly discussed
below. Dimensions are invariant under smooth transformations and thus again
computable in time delay embedding spaces.

Entropies are an information theoretical concept to characterize the amount of
information needed to predict the next measurement with a certain
precision. The most popular one is the Kolmogorov-Sinai entropy. We will
discuss here only the correlation entropy, which can be computed in a much more
robust way. The occurrence of entropies in a section on dimensions has to do
with the fact that they can be determined both by the same statistical tool.

\subsection{Correlation dimension}\label{sec:dim.c2}
Roughly speaking, the idea behind certain quantifiers of dimensions is that
the weight $p(\epsilon)$ of a typical $\epsilon$-ball covering part of the
invariant set scales with its diameter like $p(\epsilon)\approx \epsilon^D$,
where the value for $D$ depends also on the precise way one defines the
weight. Using the square of the probability $p_i$ to find a point of the set
inside the ball, the dimension is called the correlation dimension $D_2$,
which is computed most efficiently by the correlation sum~\cite{GP}:
\be\label{eq:dim2.c2}
   C(m,\epsilon)= {1\over N_{\mbox{\scriptsize pairs}}}
   \sum_{j=m}^N\sum_{k<j-w}
   \Theta(\epsilon- |\ss_j-\ss_k|)
\,,\ee
where $\ss_i$ are $m$-dimensional delay vectors, $N_{\mbox{\scriptsize
pairs}}=(N-m+1)(N-m-w+1)/2$ the number of pairs of points covered by the sums,
$\Theta$ is the Heaviside step function and $w$ will be discussed below. On
sufficiently small length scales and when the embedding dimension $m$ exceeds
the box-dimension of the attractor~\cite{SauerYorke},
\be
   C(m,\epsilon)\propto \epsilon^{D_2}
\,,\ee
Since one does not know the box-dimension {\sl a priori}, one checks for
convergence of the estimated values of $D_2$ in $m$. 

The literature on the correct and spurious estimation of the correlation
dimension is huge and this is certainly not the place to repeat all the
arguments. The relevant caveats and misconceptions are reviewed for example in
Refs.~\cite{theiler_dim,gss,dim,KantzSchreiber}. The most prominent precaution
is to exclude temporally correlated points from the pair counting by the so
called Theiler window $w$~\cite{theiler_dim}. In order to become a consistent
estimator of the correlation {\em integral} (from which the dimension is
derived) the correlation {\em sum} should cover a random sample of points drawn
independently according to the invariant measure on the attractor. Successive
elements of a time series are not usually independent. In particular for highly
sampled flow data subsequent delay vectors are highly correlated. Theiler
suggested to remove this spurious effect by simply ignoring all pairs of points
in Eq.(\ref{eq:dim2.c2}) whose time indices differ by less than $w$, where $w$
should be chosen generously. With $O(N^2)$ pairs available, the loss of $O(wN)$
pairs is not dramatic as long as $w\ll N$. At the very least, pairs with $j=k$
have to be excluded~\cite{grass_finite}, since otherwise the strong bias to
$D_2=0$, the mathematically correct value for a finite set of points, reduces
the scaling range drastically. Choosing $w$, the first zero of the
auto-correlation function, sometimes even the decay time of the autocorrelation
function, are not large enough since they reflect only overall linear
correlations~\cite{theiler_dim,dim}. The space-time-separation plot
(Sec.~\ref{sec:stp}) provides a good means of determining a sufficient value
for $w$, as discussed for example in~\cite{stp,KantzSchreiber}. In some cases,
notably processes with inverse power law spectra, inspection requires $w$ to be
of the order of the length of the time series. This indicates that the data
does not sample an invariant attractor sufficiently and the estimation of 
invariants like $D_2$ or Lyapunov exponents should be abandoned.

Parameters in the routines {\DTWO}, {\CTWO}, and {\CNAIVE} are as usual the 
embedding parameters $m$ and $\tau$, the time delay, and the embedding
dimension, as well as the Theiler window.

Fast implementation of the correlation sum have been proposed by several
authors. At small length scales, the computation of pairs can be done in
$O(N\log N)$ or even $O(N)$ time rather than $O(N^2)$ without loosing any of
the precious pairs, see Ref.~\cite{neigh}.  However, for intermediate size data
sets we also need the correlation sum at intermediate length scales where
neighbor searching becomes expensive. Many authors have tried to limit the use
of computational resources by restricting one of the sums in
Eq.(\ref{eq:dim2.c2}) to a fraction of the available points. By this practice,
however, one looses valuable statistics at the small length scales where points
are so scarce anyway that all pairs are needed for stable
results. In~\cite{buzug}, buth approaches were combined for the first time by
using fast neighbor search for $\epsilon<\epsilon_0$ and restricting the sum
for $\epsilon\ge\epsilon_0$. The {\tisean} implementations {\CTWO} and {\DTWO}
go one step further and select the range for the sums individually for each
length scale to be processed. This turns out to give a major improvement in
speed. The user can specify a desired number of pairs which seems large enough
for a stable estimation of $C(\epsilon)$, typically 1000 pairs will
suffice. Then the sums are extended to a range which guarantees that number of
pairs, or, if this cannot be achieved, to the whole time series. At the largest
length scales, this range may be rather small and the user may choose to give a
minimal number of reference points to ensure a representative average. Still,
using the program {\CTWO} the whole computation may thus at large scales be
concentrated on the first part of the time series, which seems fair for
stationary, non-intermittent data (nonstationary or strongly intermittent data
is usually unsuitable for correlation dimension estimation anyway). The program
${\DTWO}$ is safer with this respect. Rather than restricting the range of the
sums, only a randomly selected subset is used. The randomization however
requires a more sophisticated program structure in order to avoid an 
overhead in computation time.

\subsubsection{Takens-Theiler estimator}
\begin{figure}
\begin{center}
   ~\\[-12pt]
   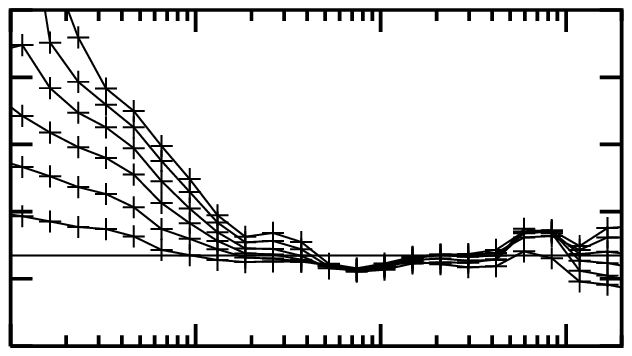\\[-7pt]
   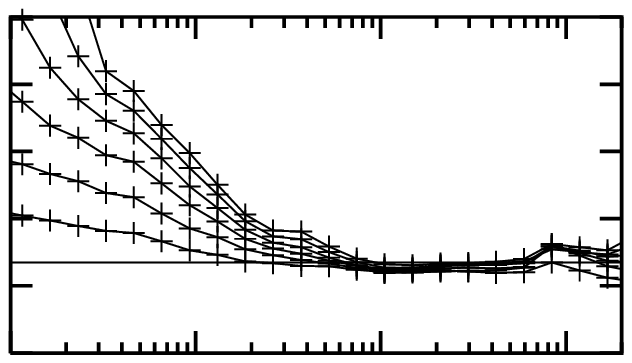\\[-7pt]
   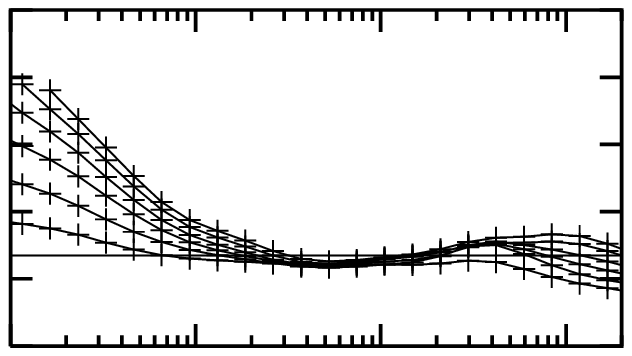\\[-7pt]
   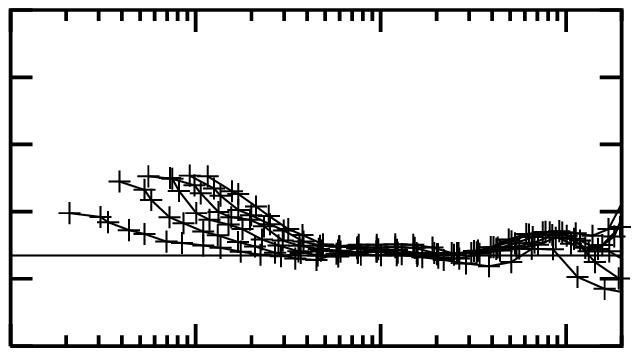
\end{center}
\caption[]{\label{fig:dim2}\small
   Dimension estimation for the (noise filtered) NMR laser data. Embedding
   dimensions 2 to 7 are shown. From above: {\bf (a)} slopes are determined by
   straight line fits to the log-log plot of the correlation sum,
   Eq.~(\ref{eq:dim2.c2}). {\bf (b)} Takes-Theiler estimator of the same slope.
   {\bf (c)} Slopes are obtained by straight line fits to the Gaussian kernel
   correlation sum, Eq.(\ref{eq:cg}). {\bf (d)} Instead of the correlation
   dimension, it has been attempted to estimate the information dimension.}
\end{figure}

Convergence to a finite correlation dimension can be checked by plotting scale
dependent ``effective dimensions'' versus length scale for various
embeddings. The easiest way to proceed is to compute (numerically) the
derivative of $\log C(m,\epsilon)$ with respect to $\log\epsilon$, for example
by fitting straight lines to the log-log plot of $C(\epsilon)$.  In
Fig.~\ref{fig:dim2}~{\bf (a)} we see the output of the routine {\CTWO} acting
on data from the NMR laser, processed by {\CTOD} in order to obtain local
slopes.  By default, straight lines are fitted over one octave in $\epsilon$,
larger ranges give smoother results. We can see that on the large scales,
self-similarity is broken due to the finite extension of the attractor, and on
small but yet statistically significant scales we see the embedding dimension
instead of a saturated, $m$-independent value. This is the effect of noise,
which is infinite dimensional, and thus fills a volume in every embedding
space. Only on the intermediate scales we see the desired {\em plateau} where
the results are in good approximation independent of $m$ and $\epsilon$. The
region where scaling is {\em established}, not just the range selected for
straight line fitting, is called the {\em scaling range}.

Since the statistical fluctuations in plots like Fig.~\ref{fig:dim2}~{\bf (a)}
show characteristic (anti-)correlations, it has been
suggested~\cite{takens_est,takens_theiler} to apply a maximum likelihood
estimator to obtain optimal values for $D_2$.  The Takens-Theiler-estimator
reads
\be\label{eq:d2t}
   D_{\mbox{\scriptsize TT}}(\epsilon)=
     {C(\epsilon) \over
     \int_0^{\epsilon}{C(\epsilon')\over\epsilon'}d\epsilon'}
\ee
and can be obtained by processing the output of {\CTWO} by {\CTOT}. Since
$C(\epsilon)$ is available only at discrete values
$\{\epsilon_i,i=0,\ldots,I\}$, we 
interpolate it by a pure power law (or, equivalently, the log-log plot by
straight lines: $\log C(\epsilon)=a_i\log\epsilon+b_i$) in between these. The 
resulting integrals can be solved
trivially and summed: 
\bes
   \int_0^{\epsilon}{C(\epsilon')\over\epsilon'}d\epsilon' &=&
   \sum_{i=1}^I e^{b_i} \int_{\epsilon_{i-1}}^{\epsilon_i} 
      (\epsilon')^{a_i-1}d\epsilon' \nonumber\\
   &=&
   \sum_{i=1}^I {e^{b_i}\over a_i} 
      (\epsilon_i^{a_i}-\epsilon_{i-1}^{a_i})
\,.\ees
Plotting $D_{\mbox{\scriptsize TT}}$ versus $\epsilon$ (Fig.~\ref{fig:dim2}
{\bf (b)}) is an interesting alternative to the usual local slopes plot,
Fig.~\ref{fig:dim2}~{\bf (a)}. It is tempting to use such an ``estimator of
dimension'' as a black box to provide a number one might quote as a dimension.
This would imply the unjustified assumption that all deviations from exact
scaling behavior is due to the statistical fluctuations. Instead, one still has
to verify the existence of a scaling regime. Only then, $D_{\mbox{\scriptsize
TT}}(\epsilon)$ evaluated at the upper end of the scaling range is a reasonable
dimension estimator.

\subsubsection{Gaussian kernel correlation integral}
The correlation sum Eq.(\ref{eq:dim2.c2}) can be regarded as an average density
of points where the local density is obtained by a kernel estimator with a step
kernel $\Theta(\epsilon-r)$. A natural modification for small point sets is to
replace the sharp step kernel by a smooth kernel function of {\em bandwidth}
$\epsilon$. A particularly attractive case that has been studied in the
literature~\cite{ghez1} is given by the Gaussian kernel, that is,
$\Theta(\epsilon-r)$ is replaced by $e^{-r^2 \over 4\epsilon^2}$.  The
resulting Gaussian kernel correlation sum $C_{\mbox{\scriptsize G}}(\epsilon)$
has the same scaling properties as the usual $C(\epsilon)$. It has been
observed in~\cite{habil} that $C_{\mbox{\scriptsize G}}(\epsilon)$ can be
obtained from $C(\epsilon)$ via
\be\label{eq:cg}
   C_{\mbox{\scriptsize G}}(\epsilon) 
      = {1\over 2\epsilon^2}\int_0^{\infty}d \tilde\epsilon \;
      e^{- {\tilde\epsilon^2\over 4\epsilon^2}} \tilde\epsilon\, 
      C(\tilde\epsilon)
\ee
without having to repeat the whole computation. If $C(\epsilon)$ is given
at discrete values of $\epsilon$, the integrals in Eq.(\ref{eq:cg}) can be
carried out numerically by interpolating $C(\epsilon)$ with pure power laws
. This is done in
{\CTOG} which uses a 15 point Gauss-Kronrod rule for the numerical integration.

\subsection{Information dimension}
Another way of attaching weight to $\epsilon$-balls, which is more natural, is
the probability $p_i$ itself. The resulting scaling exponent is called the
information dimension $D_1$. Since the Kaplan-Yorke dimension of
Sec.\ref{sec:lyap} is an approximation of $D_1$, the computation of $D_1$
through scaling properties is a relevant cross-check for highly deterministic
data. $D_1$ can be computed from a modified correlation sum, where, however,
unpleasant systematic errors occur. The {\em fixed mass}
approach~\cite{badiipoliti} circumvents these problems, so that, including
finite sample corrections~\cite{grass_finite}, a rather robust estimator
exists. Instead of counting the number of points in a ball one asks here for
the diameter $\epsilon$ which a ball must have to contain a certain number $k$
of points when a time series of length $N$ is given. Its scaling with $k$ and
$N$ yields the dimension in the limit of small length scales by
\be\label{eq:dim1.fixmass}
   D_1(m)=\lim_{k/N\to 0} {d \log k/N \over d \av{\log\epsilon(k/N)}}
\,.\ee
The routine {\CONE} computes the (geometric) mean length scale
$\exp\av{\log\epsilon(k/N)}$ for which $k$ neighbors are found in $N$ data
points, as a function of $k/N$. Unlike the correlation sum, finite sample
corrections are necessary if $k$ is small~\cite{grass_finite}. Essentially, the
$\log$ of $k$ has to be replaced by the digamma function $\Psi(k)$.  The
resulting expression is implemented in {\CONE}. Given $m$ and $\tau$, the
routine varies $k$ and $N$ such that the largest reasonable range of $k/N$ is
covered with moderate computational effort. This means that for $1/N \le k/N\le
K/N$ (default: $K=100$), all $N$ available points are searched for neighbors
and $k$ is varied. For $K/N < k/N\le 1$, $k=K$ is kept fixed and $N$ is
decreased.  The result for the NMR laser data is shown in
Fig.~\ref{fig:dim2}~{\bf (d)}, where a nice scaling with $D_1\approx 1.35$ can
be discerned. For comparability, the logarithmic derivative of $k/N$ is plotted
versus $\exp\langle\log\epsilon(k,N)\rangle$ and not vice versa, although $k/N$
is the independent variable.  One easily detects again the violations of
scaling discussed before: Cut-off on the large scales, noise on small scales,
fluctuations on even smaller scales, and a scaling range in between. In this
example, $D_1$ is close to $D_2$, and multifractality cannot be established
positively.

\subsubsection{Entropy estimates} 
The correlation dimension characterizes the $\epsilon$ dependence of the
correlation sum inside the scaling range. It is natural to ask what we can
learn form its $m$-dependence, once $m$ is larger than $D_0$. The number of
$\epsilon$-neighbors of a delay vector is an estimate of the local probability
density, and in fact it is a kind of joint probability: All $m$-components of
the neighbor have to be similar to those of the actual vector simultaneously.
Thus when increasing $m$, joint probabilities covering larger time spans get
involved. The scaling of these joint probabilities is related to the
correlation entropy $h_2$, such that
\be\label{eq:dim.fullC2}
   C(m,\epsilon) \approx \epsilon^{D_2} e^{-mh_2}
\,,\ee
As for the scaling in $\epsilon$, also the dependence on $m$ is valid
only asymptotically for large $m$, which one will not reach due to the lack
of data points. So one will study $h_2(m)$ versus $m$ and try to extrapolate to
large $m$. The correlation entropy is a lower bound of the Kolmogorov Sinai
entropy, which in turn can be estimated by the sum of the positive Lyapunov
exponents.  The program {\DTWO} produces as output the estimates of $h_2$
directly, from the other correlation sum programs it has to be extracted by
post-processing the output.

The entropies of first and second order can be derived from the output of 
{\CONE} and {\CTWO} respectively. An alternate means of obtaining these and the
other generalized entropies is by a box counting approach. Let $p_i$ be the
probability to find the system state in box $i$, then the order $q$ entropy is
defined by the limit of small box size and large $m$ of
\be\label{eq:hq}
   \sum_i p_i^q \approx e^{-mh_q}
\,.\ee   
To evaluate $\sum_i p_i^q$ over a fine mesh of boxes in $m\gg 1$ dimensions,
economical use of memory is necessary: A simple histogram would take 
$(1/\epsilon)^m$ storage. Therefore the program {\BOXCOUNT} implements
the mesh of boxes as a tree with $(1/\epsilon)$-fold branching points. The
tree is worked through recursively so that at each instance at most one
complete branch exists in storage. The current version does not implement
finite sample corrections to Eq.(\ref{eq:hq}).

\section{Testing for nonlinearity}\label{sec:surro}
Most of the methods and quantities discussed so far are most appropriate in
cases where the data show strong and consistent nonlinear deterministic
signatures. As soon as more than a small or at most moderate amount of
additive noise is present, scaling behavior will be broken and predictability
will be limited. Thus we have explored the opposite extreme, nonlinear and
fully deterministic, rather than the classical linear stochastic processes.
The bulk of real world time series falls in neither of these limiting
categories because they reflect nonlinear responses and effectively stochastic
components at the same time. Little can be done for many of these cases with
current methods. Often it will be advisable to take advantage of the
well founded machinery of spectral methods and venture into nonlinear
territory only if encouraged by positive evidence. This section is about
methods to establish statistical evidence for nonlinearity beyond a simple
rescaling in a time series.

\subsection{The concept of surrogate data}
The degree of nonlinearity can be measured in several ways. But how much
nonlinear predictability, say, is necessary to exclude more trivial
explanations? All quantifiers of nonlinearity show fluctuations but the
distributions, or error bars if you wish, are not available analytically.  It
is therefore necessary to use Monte Carlo techniques to assess the significance
of results. One important method in this context is the method of surrogate
data~\cite{theiler1}. A null hypothesis is formulated, for example that the
data has been created by a stationary Gaussian linear process, and then it is
attempted to reject this hypothesis by comparing results for the data to
appropriate realizations of the null hypothesis. Since the null assumption is
not a simple one but leaves room for free parameters, the Monte Carlo sample
has to take these into account. One approach is to construct {\em constrained
realizations} of the null hypothesis. The idea is that the free parameters left
by the null are reflected by specific properties of the data.  For example the
unknown coefficients of an autoregressive process are reflected in the
autocorrelation function. Constrained realizations are obtained by randomizing
the data subject to the constraint that an appropriate set of parameters
remains fixed. For example, random data with a given periodogram can be made by
assuming random phases and taking the inverse Fourier transform of the given
periodogram. Random data with the same distribution as a given data set can be
generated by permuting the data randomly without replacement. Asking for a
given spectrum and a given distribution at the same time poses already a much
more difficult question.

\begin{figure}[t]
\begin{center}
   ~\\[-15pt]
   \input{b.pstex}\\
   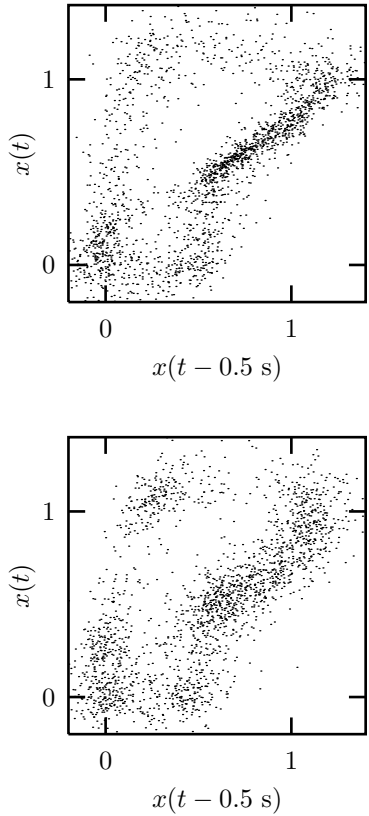
\end{center}
\caption[]{\label{fig:b_s}\small
   Upper: The human breath rate data from Fig.~\ref{fig:b}. Lower:
   the noise component extracted by the noise reduction scheme has been
   randomized in order to destroy correlations with the signal. The result 
   appears slightly but significantly less structured than the original.}
\end{figure}

\subsection{Iterative Fourier transform method}
Very few real time series which are suspected to show nonlinearity follow a
Gaussian single time distribution. Non-Gaussianity is the simplest kind of
nonlinear signature but it may have a trivial reason: The data may have been
distorted in the measurement process. Thus a possible null hypothesis would be
that there is a stationary Gaussian linear stochastic process that generates a
sequence $\{x_n\}$, but the actual observations are $s_n=s(x_n)$ where
$s(\cdot)$ is a monotonic function. Constrained realizations of this null
hypothesis would require the generation of random sequences with the same power
spectrum (fully specifying the linear process) and the same single time
distribution (specifying the effect of the measurement function) as the
observed data. The {\bf A}mplitude {\bf A}djusted {\bf F}ourier {\bf T}ransform
(AAFT) method proposed in~\cite{theiler1} attempts to invert the measurement
function $s(\cdot)$ by rescaling the data to a Gaussian distribution. Then
the Fourier phases are randomized and the rescaling is inverted. As discussed
in~\cite{surrowe}, this procedure is biased towards a flatter spectrum since
the inverse of $s(\cdot)$ is not available exactly. In the same reference,
a scheme is introduced that removes this bias by iteratively adjusting the
spectrum and the distribution of the surrogates. Alternatingly, the surrogates
are rescaled to the exact values taken by the data and then the Fourier
transform is brought to the exact amplitudes obtained from the data.
The discrepancy between both steps either converges to zero with the number
of iterations or to a finite inaccuracy which decreases with the length of the
time series. The program {\SURROGATES} performs iterations until no further
improvement can be made. The last two stages are returned, one having the exact
Fourier amplitudes and one taking on the same values as the data. For not too 
exotic data these two versions should be almost identical. The relative
discrepancy is also printed.

In Fig.~\ref{fig:b_s} we used this procedure to assess the hypothesis that 
the noise reduction on the breath data reported in Fig.~\ref{fig:b} removed 
an additive noise component which was independent of the signal. If the
hypothesis were true, we could equally well add back on the noise sequence
or a randomized version of it which lacks any correlations to the signal.
In the upper panel of Fig.~\ref{fig:b_s} we show the original data. In the
lower panel we took the noise reduced version (c.f.~Fig.~\ref{fig:b}, bottom)
and added a surrogate of the supposed noise sequence. The result is similar but
still significantly different from the original to make the additivity
assumption implausible. 

Fourier based randomization schemes suffer from some caveats due to the the
inherent assumption that the data constitutes one period of a periodic signal,
which is not what we really expect. The possible artefacts are discussed for
example in~\cite{theiler_sfi} and can, in summary, lead to spurious rejection
of the null hypothesis. One precaution that should be taken when using
{\SURROGATES} is to make sure that the beginning and the end of the data
approximately match in value and phase. Then, the periodicity assumption is not
too far wrong and harmless. Usually, this amounts to the loss of a few points
of the series. One should note, however, that the routine may truncate the data
by a few points itself in order to be able to perform a {\em fast} Fourier
transform which requires the number of points to be factorizable by small prime
factors.

\begin{figure}
~\\[-5pt]
\centerline{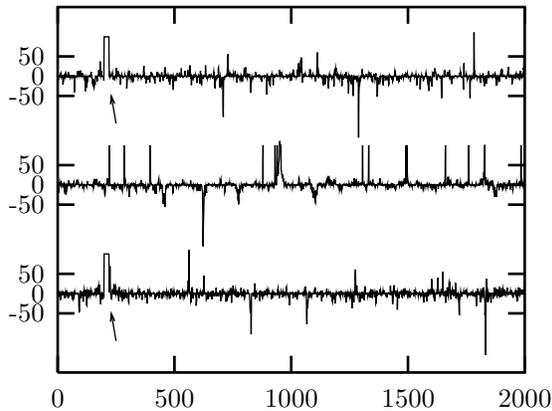}
\caption[]{\label{fig:spike}\small
   Upper trace: Data from a stationary Gaussian linear stochastic process
   ($x_n=0.7x_{n-1}+\eta_n$) 
   measured by $s(x_n)=x_n^3$. Samples 200-220 are an artefact. With the
   Fourier based scheme (middle trace) the artefact results in an increased
   number of spikes in the surrogates and reduced predictability. In the lower
   trace, the artefact has been preserved along with the distribution of values
   and lags $1,\ldots,25$ of the autocorrelation function.}
\end{figure}

\begin{figure}[t]
\begin{center}
   ~\\[-15pt]
   \mbox{\hspace*{-0.5cm}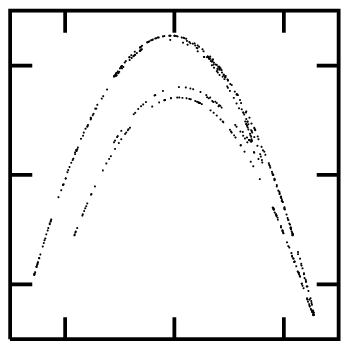%
   \hspace*{-0.9cm}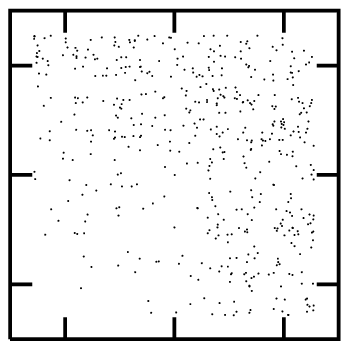}\\[-10pt]
   \mbox{\hspace*{-0.5cm}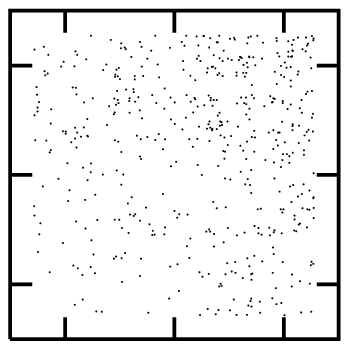%
   \hspace*{-0.9cm}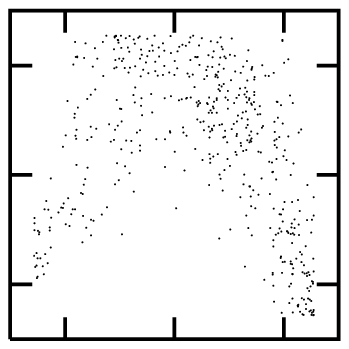}\\[-10pt]
   \mbox{\hspace*{-0.5cm}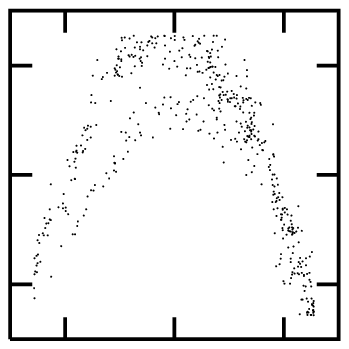%
   \hspace*{-0.9cm}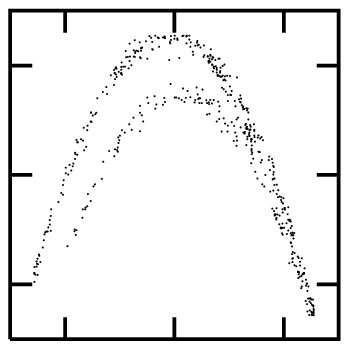}
\end{center}
\caption[]{\label{fig:henon}\small
  Randomization of 500 points generated by the the H\'enon map.  {\bf (a)}
  Original data; {\bf (b)} Same autocorrelations and distribution; {\bf
  (c)}-{\bf (f)} Different stages of annealing with a cost function $C$
  involving three and four-point correlations. {\bf (c)} A random shuffle,
  $C=2400$; {\bf (d)} $C=150$; {\bf (e)} $C=15$; {\bf (f)} $C=0.002$. See
  text.}
\end{figure}

\subsection{General constrained randomization}
In~\cite{anneal}, a general method has been proposed to create random data
which fulfill specified constraints. With this method, the artefacts and
remaining imprecision of the Fourier based randomization schemes can be
avoided by specifying the autocorrelation function rather than the Fourier
transform. The former does not assume periodic continuation. Maybe more
importantly, the restriction to a rather narrow null hypothesis can be relaxed
since in principle arbitrary statistical observables can be imposed on the
surrogates. A desired property of the data has to be formulated in terms of a
cost function which assumes an absolute minimum when the property is fulfilled.
States arbitrarily close to this minimal cost can be reached by the method of
simulated annealing. The cost function is minimised among all possible
permutations of the data. See~\cite{anneal} for a description of the approach.

The {\tisean} package contains the building blocks for a library of surrogate
data routines implementing user specified cost functions. Currently, only the
autocorrelation function with and without periodic continuation have been
implemented. Further, a template is given from which the user may
derive her/his 
own routines. A module is provided that drives the simulated annealing process
through an exponential cooling scheme. The user may replace this module by 
other scheme of her/his choice. 
A module that performs random pair permutations is
given which allows to exclude a list of points from the permutation scheme.
More sophisticated permutation schemes can be substituted if desired.
Most importantly, the cost function has to be given as another module. 
The autocorrelation modules use $\max_{\tau=1}^{\tau_{\mbox{\scriptsize max}}}
 | C(\tau)-C(\tau)_{\mbox{\scriptsize data}} |/\tau$, where $C(\tau)$ is the
autocorrelation function with or without periodic continuation.

In Fig.~\ref{fig:spike} we show an example fulfilling the null hypothesis of a
rescaled stationary Gaussian linear stochastic process which has been
contaminated by an artefact at samples 200-220. The Fourier based schemes are
unable to implement the artefact part of the null hypothesis. They spread the
structure given by the artefact evenly over the whole time span, resulting in
more spikes and less predictability. In fact, the null hypothesis of a
stationary rescaled Gaussian linear stochastic process can be rejected at the
95\% level of significance using nonlinear prediction errors. The artefact
would spuriously be mistaken for nonlinearity. With the program {\RAUTO},
we can exclude the artefact from the randomization scheme and obtain a correct
test. 

As an example of a more exotic cost function, let us show the randomization 
of 500 iterates of the H\'enon map, Fig.~\ref{fig:henon}~{\bf (a)}. Panel {\bf
(b)} shows the output of {\SURROGATES} having the same spectrum and
distribution. Starting from a random permutation {\bf (c)}, the cost function
\bes
   C &=& \av{x_{n-1}x_n} + \av{x_{n-2}x_n} \nonumber\\
     &+& \av{x_{n-1}^2x_n} + \av{x_{n-1}x_n^2} +
          \av{x_{n-2}^2x_n}+\av{x_{n-2}x_{n-1}x_n} \nonumber\\
     &+& \av{x_{n-1}^2x_n^2} +\av{x_{n-1}x_n^3} +\av{x_{n-1}^3x_n}
\ees
is minimized ({\RGENER}). It involves are all the higher order
autocorrelations which would be needed for a least squares fit with the ansatz
$x_n=c-ax_{n-1}^2+bx_{n-2}$ and in this sense fully specifies the quadratic
structure of the data. The random shuffle yields $C=2400$, panels 
{\bf (c)}-{\bf (f)} correspond to $C=150,15,0.002$ respectively.

Since the annealing process can be very CPU time consuming, it is important to
provide efficient code for the cost function.  Specifying
$\tau_{\mbox{\scriptsize max}}$ lags for $N$ data points requires
$O(N\tau_{\mbox{\scriptsize max}})$ multiplications for the calculation of the
cost function. An update after a pair has been exchanged, however, can be
obtained with $O(\tau_{\mbox{\scriptsize max}})$ multiplications. Often, the
full sum or supremum can be truncated since after the first terms it
is clear that a large increase of the cost is unavoidable. The driving
Metropolis algorithm provides the current maximal permissable cost for that
purpose.

The computation time required to reach the desired accuracy depends on the
choice and implementation of the cost function but also critically on the
annealing schedule. There is a vast literature on simulated annealing which
cannot be reviewed here. Experimentation with cooling schemes should keep in
mind the basic concept of simulated annealing. At each stage, the system --
here the surrogate to be created -- is kept at a certain ``temperature''.
Like in thermodynamics, the temperature determines how likely fluctuations
around the mean energy -- here the value of the cost function $C$ -- are.  At
temperature $T$, a deviation of size $\Delta C$ occurs with the Boltzmann
probability $\propto \exp(-\Delta C/T)$. In a Metropolis simulation, this is
achieved by accepting {\em all} downhill changes ($\Delta C<0$), but also
uphill changes with probability $\exp(-\Delta C/T)$. Here the changes are
permutations of two randomly selected data items. The present implementation
offers an exponential cooling scheme, that is, the temperature is lowered by a
fixed factor whenever one of two conditions is fulfilled: Either a specified
number of changes has been {\em tried}, or a specified number of changes has
been {\em accepted}. Both these numbers and the cooling factor can be chosen
by the user. If the state is cooled too fast it gets stuck, or ``freezes'' in
a false minimum. When this happens, the system must be ``melted'' again and
cooling is taken up at a slower rate. This can be done automatically until a
goal accuracy is reached. It is, however, difficult to predict how many steps
it will take. The detailed behavior of the scheme is still subject to ongoing
research and in all but the simplest cases, experimentation by the user will
be necessary. To facilitate the supervision of the cooling, the current state
is written to a file whenever a substantial improvement has been
made. Further, the verbosity of the diagnostic output can be selected.

\subsection{Measuring weak nonlinearity}
When testing for nonlinearity, we would like to use quantifiers which are
optimized for the weak nonlinearity limit, which is not what most time series
methods of chaos theory have been designed for. The simple nonlinear prediction
scheme (Sec.~\ref{sec:zeroth}) has proven quite useful in this context. If used
as a comparative statistic, it should be noted that sometimes seemingly
inadequate embeddings or neighborhood sizes may lead to rather big errors
which have, however, small fluctuations. The tradeoff between bias and variance
may be different from the situation where predictions are desired {\em per
se}. The same rationale applies to quantities derived from the correlation
sum. Neither the small scale limit, genuine scaling, or the Theiler correction
are formally necessary in a comparative test. However, any temptation to
interpret the results in terms like ``complexity'' or ``dimensionality'' should
be resisted, even though ``complexity'' doesn't seem to have an agreed-upon
meaning anyway. Apart from average prdiction errors, we have found the
stabilities of short periodic orbits (see Sec.~\ref{sec:upo}) useful for the
detectionof nonlinearity in surrogate data tests. As an alternative to the
phase space based methods, more traditional measures of nonlinearity derived
from higher order autocorrelation functions (\cite{BI}, routine
{\AUTOCORTHREE}) may also be considered. If a time-reversal asymmetry is
present, its statistical confirmation (routine {\TIMEREV}) is a very powerful
detector of nonlinearity~\cite{diks2}. Some measures of weak nonlinearity are
compared systematically in Ref.~\cite{power}.

\section{Conclusion and perspectives}\label{sec:conclude}
The {\tisean} project makes available a number of algorithms of nonlinear time
series analysis to people interested in applications of the dynamical systems
approach. To make proper use of these algorithms, it is not essential to have
witten the programs from scratch, an effort we intend to spare the user by
making {\tisean} public. Indispensable, however, is a good knowledge of what
the programs do, and why they do what they do. The latter requires a thorough
background in the nonlinear time series approach which cannot be provided by
this paper but rather by textbooks like Refs.~\cite{abarbook,KantzSchreiber},
reviews~\cite{gss,abarbanel,habil}, and the original literature~\cite{coping}.
Here, we have concentrated on the actual implementation as it is realized in
{\tisean} and on examples of the concrete use of the programs.

\subsection{Important methods which are (still) missing}
Let us finish the discussion by giving some perspectives on future work.
So far, the {\tisean} project has concentrated on the most common situation
of a single time series. While for multiple measurements of similar nature
most programs can be modified with moderate effort, a general framework
for heterogeneous multivariate recordings (say, blood pressure and heart beat)
has not been established so far in a nonlinear context. Nevertheless, we feel
that concepts like generalized synchrony, coherence, or information flow
are well worth pursuing and at some point should become available to a wider
community, including applied research.

Initial experience with nonlinear time series methods indicates that some of
the concepts may prove useful enough in the future to become part of the
established time series tool box. For this to happen, availability of the
algorithms and reliable information on their use will be essential.  The
publication of a substantial collection of research level programs through the
{\tisean} project may be seen as one step in that direction.  However, the
potential user will still need considerable experience in order to make the
right decisions -- about the suitability of a particular method for a specific
time series, about the selection of parameters, about the interpretation of
the results. To some extent, these decisions could be guided by software that
evaluates the data situation and the results automatically. Previous
experience with black box dimension or Lyapunov estimators has not been
encouraging, but for some specific problems, ``optimal'' answers can in
principle be defined and computed automatically, once the optimality criterion
is formulated. For example, the prediction programs could be encapsulated in a
framework that automatically evaluates the performance for a range of
embedding parameters etc. Of course, quantitative assessment of the results is
not always easy to implement and depends on the purpose of the study.
As another example, it seems realistic to define ``optimal'' Poincar\'e
surfaces of section and to find the optimal solutions numerically.

Like in most of the time series literature, the issue of stationarity has
entered the discussion only as something the lack of which has to be detected
in order to avoid spurious results. Taking this point seriously amounts to
rejecting a substantial fraction of time series problems, including the most
prominent examples, that is, most data from finance, metereology, and biology.
It is quite clear that the mere rejection of these challenging problems is not
satisfactory and we will have to develop tools to actually analyse, understand,
and predict nonstationary data. Some suggestions have been made for the
detection of fluctuating control parameters~\cite{Kadtke,B1,casEEG,statio}.
Most of these can be seen as continuous versions of the classification problem,
another application which is not properly represented in {\tisean} yet.

Publishing software, or reviews and textbooks for that matter, in a field
evolving as rapidly as nonlinear time series analysis will always have the
character of a snapshot of the state at a given time. Having the options either
to wait until the field has saturated sufficiently or to risk that programs, or
statements made, will become obsolete soon, we chose the second option.
We hope that we can thus contibute to the further evolution of the field.

\section*{Acknowledgments}
We wish to thank Eckehard Olbrich, Marcus Richter, and Andreas Schmitz who have
made contributions to the {\tisean} project, and the users who patiently coped
with early versions of the software, in particular Ulrich Hermes.  
We thank Leci Flepp, Nick Tufillaro,
Riccardo Meucci, and Marco Ciofini for letting us use their time series data.
This work was supported by the SFB 237 of the Deutsche Forschungsgemeinschaft.


\begin{thebibliography}{}
\bibitem{tisean}
   The {\tisean} software package is publicly available at \THTTP.
   The distribution includes an online documentation system.
\end{thebibliography}
\end{document}